\documentclass[lettersize,journal]{IEEEtran}

\usepackage{cite}

\usepackage{stfloats}

\usepackage{amsmath,amsfonts}
\usepackage[ruled,boxed,linesnumbered]{algorithm2e}

\usepackage{listings}
\usepackage{graphicx}
\usepackage{subfigure}

\usepackage{multirow}
\usepackage{makecell}
\usepackage{threeparttable}

\usepackage{stfloats}

\usepackage{textcomp}
\usepackage{xcolor}

\usepackage{verbatim}

\usepackage[colorlinks=true,linkcolor=blue,citecolor=blue,urlcolor=blue]{hyperref}

\def\BibTeX{{\rm B\kern-.05em{\sc i\kern-.025em b}\kern-.08em
		T\kern-.1667em\lower.7ex\hbox{E}\kern-.125emX}}
	
\begin{document}

\title{FASE: FPGA-Assisted Syscall Emulation for Rapid Early-Stage Processor Performance Evaluation}

\author{
	\IEEEauthorblockN{Chengzhen Meng\IEEEauthorrefmark{1},
		Xiuzhuang Chen\IEEEauthorrefmark{2},
		Bingcai Sui\IEEEauthorrefmark{3}\IEEEauthorrefmark{4},
		Zhenyu Zhao\IEEEauthorrefmark{3}\IEEEauthorrefmark{4},
		Tun Li\IEEEauthorrefmark{4},
		Hongjun Dai\IEEEauthorrefmark{1}\thanks{Corresponding author: Hongjun Dai (email: dahogn@sdu.edu.cn)}
	}
	
	\IEEEauthorblockA{\IEEEauthorrefmark{1} School of Software, Shandong University, Jinan, China}
	
	\IEEEauthorblockA{\IEEEauthorrefmark{2} School of Integrated Circuits, Shandong University, Jinan, China}
	
	\IEEEauthorblockA{\IEEEauthorrefmark{3} Key Laboratory of Advanced Microprocessor Chips and Systems, National University of Defense Technology, Changsha, China}
	
	\IEEEauthorblockA{\IEEEauthorrefmark{4} College of Computer Science and Technology, National University of Defense Technology, Changsha, China}
}

\maketitle

\begin{abstract}
	
	The rapid advancement of AI workloads and domain-specific architectures has led to increasingly diverse processor microarchitectures, whose design exploration requires fast and accurate performance validation.
	However, traditional workflows defer validation process until RTL design and SoC integration are complete, significantly prolonging development and iteration cycle.
	
	In this work, we present FASE framework, FPGA-Assisted Syscall Emulation, the first work for adapt syscall emulation on FPGA platforms, enabling complex multi-thread benchmarks to directly run on the processor design without integrating SoC or target OS for early-stage performance evaluation.
	FASE introduces three key innovations to address three critical challenges for adapting FPGA-based syscall emulation:
	(1) only a minimal CPU interface is exposed, with other hardware components untouched, addressing the lack of a unified hardware interface in FPGA systems;
	(2) a Host-Target Protocol (HTP) is proposed to minimize cross-device data traffic, mitigating the low-bandwidth and high-latency communication between FPGA and host;
	and (3) a host-side runtime is proposed to remotely handle Linux-style system calls, addressing the challenge of cross-device syscall delegation.
	
	Experiments were conducted on Xilinx FPGA with open-sourced RISC-V SMP processor Rocket.
	With single-thread CoreMark, FASE introduces less than 1\% performance error and achieves over 2000× higher efficiency compared to Proxy Kernel due to FPGA acceleration.
	With complex OpenMP benchmarks, FASE demonstrates over 96\% performance validation accuracy for most single-thread workloads and over 91.5\% for most multi-thread workloads compared to full SoC validation, significantly reducing development complexity and time-to-feedback.
	All components of FASE framework are released as open-source \footnote{https://github.com/meng-cz/fase-rv64}.
	
\end{abstract}


\section{Introduction}

	
	In today’s data-driven world, the evolution of AI computing instruction sets and domain-specific architectures has accelerated processor innovation.
	As a result, fast and accurate performance evaluation is required to provide quantitative evidence for agile development and processor design optimization \cite{ref-byg, ref-modse-exploration, ref-presi}.
	In CPU design, early-stage performance evaluation is particularly critical. It directly influences microarchitecture decisions, reduces the risk of costly redesign, and accelerates the development of emerging architectures such as matrix ISAs and chiplet-based systems \cite{ref-rv-isa-extend, ref-rvcnn, ref-cad-assisted-sim1}.

\begin{figure} [t]
		\centering
		\subfigure[In the traditional workflow, full hardware-software integration on electronic design automation (EDA) tools is required before processor evaluation.]{
			\begin{minipage}{\columnwidth}
				\centering
				\includegraphics[width=\linewidth]{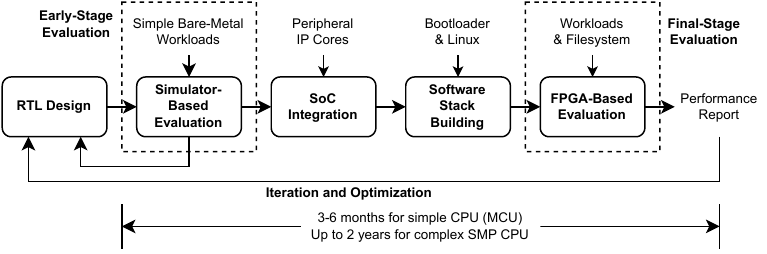}
			\end{minipage}
		}
		\\
		\subfigure[Rapid end-to-end performance evaluation approached enable direct loading and execution of the final user-state workloads on the processor design.]{
			\begin{minipage}{\columnwidth}
				\centering
				\includegraphics[width=0.8\linewidth]{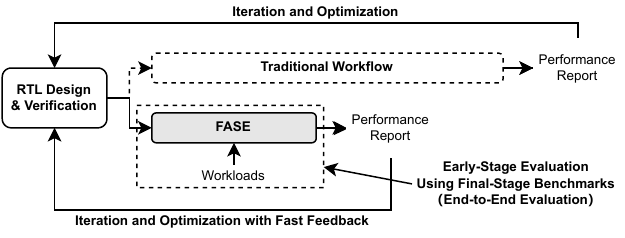}
			\end{minipage}
			\label{fig-intro2}
		}
		\caption{Utilizing FASE on FPGA to streamline the performance evaluation workflow in the iteration process of novel processor designs.}
		\label{fig-intro}
	\end{figure}

	However, time-efficient and accurate early-stage CPU performance evaluation remains an open problem.
	Common techniques include analytical performance models (e.g., the roofline model) \cite{ref-roofline-e2e, ref-roofline-gpu} and system simulation frameworks \cite{ref-survey-cachesim, ref-bzsim}.
	Analytical models provide only coarse estimates, while simulators suffer from trade-offs among speed, accuracy, and engineering effort.
	Following common practice, the terms \textbf{accuracy} and \textbf{fidelity} denote the degree to which evaluation results match the performance of a fabricated chip running the full software stack.
	As illustrated in Figure \ref{fig-intro}, current CPU design flows typically divide performance evaluation into two phases: (1) early-stage evaluation on software simulators; and (2) final-stage evaluation on FPGA prototypes.
	Early-stage performance insights often have limited fidelity, because long simulation time and incomplete target system restrict the workloads to simplified or partial programs.
	Final-stage evaluation can execute full benchmark suites with near-final fidelity, but it requires full system integration and substantial engineering effort.
	The time gap between these two phases often spans several months \cite{ref-sim-techs}.
	Therefore, the ability to run full benchmark workloads during early-stage evaluation with both high time-efficiency and accuracy would significantly accelerate design iteration \cite{ref-end2end-sim-dlworkload, ref-agile-test}.
	
	Existing studies in early-stage processor performance evaluation primarily focused on hardware modeling and simulation techniques, seeking to mitigate the efficiency–accuracy trade-off \cite{ref-gem5-hetosoc-sim, ref-boom-sim, ref-manycore-framework, ref-esesc}.
	In contrast, we takes a fundamentally different new perspective: instead of refining simulators, we target the key barrier that prevents early-stage processor prototypes from running benchmarks with FPGA acceleration.
	
	This barrier originates from the fact that complex benchmarks are user programs that consist of only user-mode instruction and depend on system calls (syscalls) for OS services such as file I/O and memory management \cite{ref-gem5}.
	Consequently, traditional FPGA-based pre-silicon evaluations must incorporate a complete hardware–software system to provide the necessary syscall environment for benchmarks \cite{ref-chipyard}.
	
	Therefore, our motivation is to leverage host resources to provide syscall services for FPGA-based target systems.
	In this way, as long as the target hardware can correctly execute user-mode instructions, it is capable of running complex user programs directly on FPGA.
	Furthermore, if thread-related syscalls are supported, performance evaluation of multicore prototypes can also benefit from FPGA acceleration \cite{ref-mcsima, ref-sniper}.
	
	This shift brings two major advantages for agile processor design:
	\begin{itemize}
		\item Accurate performance evaluation on FPGA can be carried out much earlier in the design cycle, with application scenarios covering from an initial functional pipeline prototype to more advanced multi-core designs.
		\item This avoids the engineering overhead of integrating peripheral IPs and porting software stacks during microarchitecture exploration, allowing designers to focus on innovation while still gaining representative performance insights.
	\end{itemize}

	However, although similar ideas have been widely adopted in software-based microarchitecture simulators, known as Syscall Emulation \cite{ref-rvvec-sim-bench-nogem5, ref-zsim, ref-mcsima}, no prior work has been reported on applying Syscall Emulation to FPGA-based systems.
	Three key challenges underlie this gap:
	First, unlike software simulators that operate on well-defined hardware models (e.g., GEM5 with the Ruby memory system), FPGA-based systems lack a unified interface for accessing or controlling target hardware during remote syscall handling.
	Second, the communication bandwidth between host and FPGA is inherently limited, which increases exception handling latency and impair performance evaluation accuracy.
	Third, multi-threaded system calls requires tracking fine-grained synchronization between CPU cores, while relevant hardware circuits are invisible to the host.
	Section \ref{section-backgrounds} provides a detailed introduction to the fundamentals of Syscall Emulation and the challenges of implementing it on FPGAs.
	
	In this paper, we present FASE (FPGA-Assisted Syscall Emulation), a framework that addresses these three challenges through a set of novel techniques spanning the target-hardware, communication, and host-software layers.
	FASE requires only a minimal CPU debug interface to be exposed, without any modifications to the rest of target hardware, ensuring broad applicability across processor designs.
	FASE introduces a new Host-Target Protocol (HTP) and a dedicated hardware controller to minimize data transmission over low-bandwidth UART channels during syscall emulation.
	FASE integrates a host-side runtime equipped with I/O redirection, remote thread scheduling and synchronization, and target virtual memory management.
	Section \ref{section-overview} to \ref{section-host} describe the FASE framework and the proposed methods in detail.
	
	We validated FASE with the open-source 64-bit RISC-V processor Rocket and deploy it on a Xilinx KCU105 FPGA. 
	FASE successfully executed dynamically linked, multi-threaded user applications on a target system consisting of only the CPU cores and DDR.
	On simple workload CoreMark, FASE achieves over 99\% performance evaluation accuracy and achieves over 2000× higher efficiency compared to Proxy Kernel through FPGA acceleration.
	On complex open-source benchmarks, FASE achieves over 96\% performance evaluation accuracy on single-thread workloads, and over 91.5\% accuracy in most multi-thread workloads.
	Section \ref{section-exp} presents detailed experimental results and error analysis, while Section \ref{section-discussion} discusses current challenges and future work for FASE.
	
	The main contributions of this paper are as follows:
	\begin{itemize}
		\item We propose FASE, the first work to enable syscall emulation on FPGA prototypes, directly addressing the long-standing bottleneck in which complex benchmarks cannot be executed until full-system integration is completed.
		\item We implement FASE on a RISC-V 64-bit ISA, with all source code released as open source.
		\item We experimentally demonstrate FASE’s correctness and performance fidelity for multi-threaded benchmarks, ensuring that its evaluation results remain meaningful and actionable in practical workflow.
		\item We introduce a new agile processor development paradigm by enabling early performance feedback, reducing the iteration time and supporting broader design-space exploration on RTL-modeled hardware.
	\end{itemize}
	
\section{Backgrounds}
\label{section-backgrounds}

\subsection{Syscall Emulation}
	

	We illustrate syscall emulation with a simple matrix multiplication C program using \emph{OpenBLAS} and \emph{glibc}.
	Computational kernels (e.g., \emph{dgemm}) execute entirely in user mode, while I/O operations such as \emph{printf} eventually invoke syscalls (e.g., \emph{write}) to request OS services.
	Upon a syscall instruction, control transfers to the OS, which performs privileged operations (e.g., file validation and device interaction) before returning to user space.
	

	


	\begin{figure} [t]
		\centering
		\subfigure[In a full Linux system, system calls are handled by privileged code.]{
			\begin{minipage}{\columnwidth}
				\centering
				\includegraphics[width=0.7\linewidth]{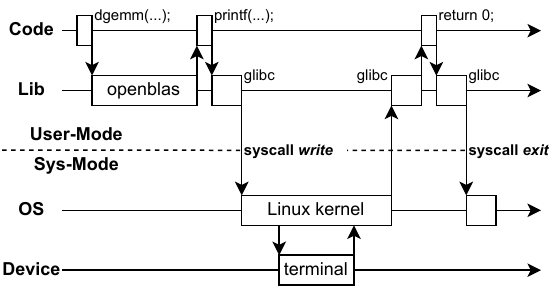}
			\end{minipage}
			\label{fig-fs-example}
		}
		\subfigure[In FASE, the CPU is emulated on FPGA while system calls are remotely handled on host side.]{
			\begin{minipage}{\columnwidth}
				\centering
				\includegraphics[width=0.7\linewidth]{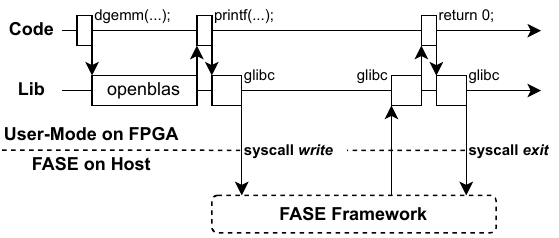}
			\end{minipage}
			\label{fig-se-example}
		}
		\caption{Example of system call handling in full Linux and syscall emulation.}
		\label{fig-example}	
	\end{figure}

	Therefore, two properties enable syscall emulation:
	\begin{itemize}
		\item[1.] Syscalls are the standardized interface between user programs and the kernel.
		\item[2.] Only OS-dependent operations (I/O, process and memory management) require syscalls, whereas most computation remains in user mode.
	\end{itemize}

	Syscall emulation leverages this interface abstraction by reproducing the Linux syscall contract without executing kernel code, as shown in Figure \ref{fig-se-example}.
	User programs observe correct behavior as long as syscall arguments, return values, and architectural state updates conform to the Linux specification.
	Software simulators such as QEMU and GEM5 implement syscall handling for this purpose.
	In contrast, FASE separates execution and syscall handling across devices, introducing cross-device coordination challenges.

	For multi-threaded workloads, most synchronization occurs in user space via atomic operations, while the kernel is involved only for blocking primitives (e.g., \emph{futex}).
	Therefore, syscall emulation primarily affects blocking and I/O behavior rather than regular computation.


	
	
	The correctness of syscall emulation depends on: (1) faithful execution of user-mode instructions, and (2) Linux-compatible syscall semantics.
	Its performance accuracy is thus determined by the fraction of user-mode execution, making it suitable for compute-intensive workloads but less precise for I/O- or synchronization-dominated applications.

\subsection{Challenges for Porting Syscall Emulation to FPGA}

	
	

	Porting syscall emulation to FPGA introduces additional challenges.
	Syscalls require privilege transitions, architectural state access, and MMU interaction, yet FPGA-based hardware lacks the unified control mechanisms available in software simulators.
	Moreover, syscall arguments and results must traverse host–FPGA boundaries, making bandwidth and latency critical.
	Multi-threaded syscalls further introduce coordination overhead and require more complex mechanisms.

\section{Overview of FASE Framework}
\label{section-overview}

	\begin{figure} [t]
		\centering
		\includegraphics[width=0.9\linewidth]{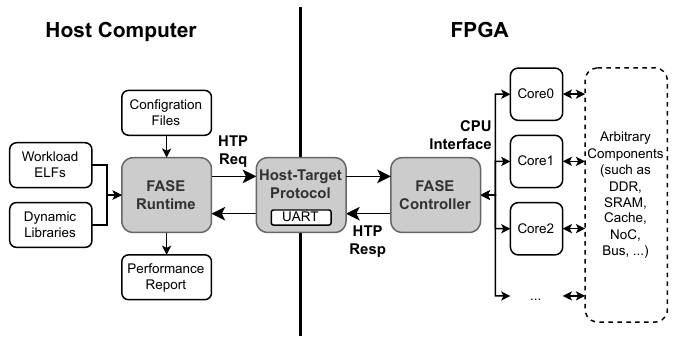}
		\caption{The architecture overview of the proposed syscall emulation framework on FPGA. The shaded components are introduced by our method.}
		\label{fig-overview}
	\end{figure}

	Figure \ref{fig-overview} illustrates the overall architecture of the proposed FASE system, which enables syscall emulation on FPGA-based hardware platforms.
	The shaded regions in the diagram represent the components introduced by our method, encompassing both the on-chip hardware and the host-side software.
	
	FASE is designed to simulate multicore processors and memory/cache systems on FPGA without modeling peripheral devices or deploying a full operating system.
	Instead, system calls and exceptions generated by the emulated user program are delegated to the host computer through a UART interface.
	The entire implementation, including hardware modules and host-side software, has been released as open-source to facilitate adoption and further development.
	
	At the hardware level, FASE introduces a controller module connected to the processor pipeline to manage the core and communicate with the host.
	To minimize porting effort to new designs, the target core only needs to expose three signal groups: privilege state, register access, and non-branch instruction injection.
	Also, the FASE Host-Target Protocol (HTP) is introduced to consolidate common architecture-level operations and reduce cross-device communication overhead.
	Section \ref{section-hardware} describes the specific core-level interfaces required for integrating the FASE mechanisms and elaborates on the internal structure of the FASE Controller and HTP.
	
	At the software level, a sophisticated FASE runtime system on the host is designed to provide remote handling of thread management, virtual memory, and I/O system calls.
	Users can execute workloads on FASE-enabled processors in FPGA by simply providing ELF binaries, dynamic libraries if needed, and configuration files on the host.
	FASE will also profile the execution and collect performance reports for the target workload.
	Section \ref{section-host} outlines the design and implementation of the host-side software responsible for workload management and syscall handling.

\section{FASE CPU Interface and Hardware Controller}	
\label{section-hardware}

	This section presents the design of the FASE hardware framework, consisting of a FASE Controller together with corresponding target CPU interface and host communication mechanism.
	The FASE Controller is released as a parameterizable module within the open-source FASE code.
	
	The principal challenge in designing the FASE Controller lies in two point: (1) how to realize all necessary interactions for remote exception processing through a minimal set of hardware ports; (2) how to design an optimized communication intermediary that minimizes cross-device data traffic.
	
	To this end, a streamlined CPU debug interface is adopted as the target hardware interface to maximize reuse and minimize porting effort, and the \emph{FASE Host-Target Protocol (HTP)} is introduced to reduce unnecessary data transfers under low-bandwidth, high-latency cross-device communication conditions.
	As the core of the hardware framework, the FASE Controller handles HTP requests from the host and manages the execution of target processors via the CPU interface.
	
\subsection{CPU Control Interface for Syscall Handling}
\label{section-interface-ports}

	This section introduces the target hardware interface definition required to support the FASE framework.
	Notably, this constitutes the only necessary modification to the target processor when applying our approach.
	All other components of the framework, including the hardware controller and host-side software, can operate independently of the processor’s internal implementation.
	Each logical CPU core should expose a separate CPU interface.

	\begin{table}[t]
		\centering
		\caption{Required CPU Ports Defined in FASE}
		\resizebox{0.9\linewidth}{!}{
		\begin{tabular}{|c|c|c|c|}
			\hline
			\textbf{Bundle}         & \textbf{Port}                                                 & \textbf{Direction}                    & \textbf{Description / Operation}                                                                                        \\ \hline
			Priv                    & Priv                                                          & CPU \textgreater FASE                  & Current hardware privilege level                                                                                        \\ \hline
			\multirow{4}{*}{Reg}    & \begin{tabular}[c]{@{}c@{}}RegAccess\\ (vld-rdy)\end{tabular} & \multirow{3}{*}{FASE \textgreater CPU} & \begin{tabular}[c]{@{}c@{}}Read or write a general purpose register\\ (valid-ready handshake signal group)\end{tabular} \\ \cline{2-2} \cline{4-4} 
			& RegWEN                                                        &                                       & Write enable                                                                                                            \\ \cline{2-2} \cline{4-4} 
			& RegIdx                                                        &                                       & Architecture register index                                                                                             \\ \cline{2-4} 
			& RegData                                                       & CPU \textless{}\textgreater FASE       & Read/Write data                                                                                                         \\ \hline
			\multirow{4}{*}{Inject} & StopFetch                                                     & \multirow{3}{*}{FASE \textgreater CPU} & Clutch fetch unit and decoder unit                                                                                      \\ \cline{2-2} \cline{4-4} 
			& \begin{tabular}[c]{@{}c@{}}Inject\\ (vld-rdy)\end{tabular}    &                                       & \begin{tabular}[c]{@{}c@{}}Inject a non-branch instruction\\ (valid-ready handshake signal group)\end{tabular}         \\ \cline{2-2} \cline{4-4} 
			& InjectInst                                                    &                                       & Raw instruction to be injected                                                                                           \\ \cline{2-4} 
			& InjectBusy                                                    & CPU \textgreater FASE                  & Execution pipeline not empty                                                                                            \\ \hline
			\textit{Optional}                    & Interrupt                                                          & FASE \textgreater CPU                  & Raise a hardware interrupt                                                                                        \\ \hline
		\end{tabular}
	}
	\label{tab-cpu-port}
	\end{table}
	
	After extensive design-space exploration and empirical validation, the required interface was reduced to three bundles.
	Table \ref{tab-cpu-port} summarizes the proposed CPU interface, which is organized into three bundles: \emph{Priv}, \emph{Reg}, and \emph{Inject}.
	The \emph{Priv} bundle reports the current hardware privilege level for exception detection.
	The \emph{Reg} bundle provides controlled access to architectural registers through a standard handshake mechanism.
	The \emph{Inject} bundle enables back-end instruction injection, including front-end isolation, injection handshake, and pipeline status reporting.
	An optional \emph{interrupt} signal supports more advanced scheduling and synchronization, reflecting a deliberate trade-off to limit hardware
	integration complexity.
	
	These capabilities are fundamental and already covered by widely adopted debug standards such as the \emph{RISC-V Debug Specification}.
	
	In practice, redundant privileged instruction execution is prevented through two complementary mechanisms: (1) \emph{StopFetch} signal is invalid only during user program execution; and (2) the interrupt vector is redirected to a simple infinite loop.

\subsection{FASE Host-Target Protocol (HTP)}
\label{section-controller-protocol}

	\begin{table}[t]
		\centering
		\caption{HTP Requests and Corresponding Execution Patterns}
		\resizebox{\linewidth}{!}{
		\begin{threeparttable}
		\begin{tabular}{|c|c|l|}
			\hline
			\textbf{Group}                                                                 & \textbf{Request} & \textbf{Execution Pattern over CPU Ports \textsuperscript{1} (use RISC-V as examples)}                                                                                                                                              \\ \hline
			\multirow{5}{*}{\begin{tabular}[c]{@{}c@{}}Inst-\\ Stream\\ Control\end{tabular}} & Redirect         & x1=\textsuperscript{2}addr; \$\textsuperscript{3}csrs mstatus,3\textless{}\textless{}11; \$csrw mepc,x1; \$mret;                                                                                                                                   \\ \cline{2-3} 
			& Next             & \begin{tabular}[c]{@{}l@{}}wait exception\textsuperscript{4}; \$csrr x1,mcause; \$csrr x2,mepc; \$csrr x3,mtval; \\  send\textsuperscript{5} x1,x2,x3; \end{tabular}                                                                                                       \\ \cline{2-3} 
			& MMU              & \begin{tabular}[c]{@{}l@{}}(1) SetMMU: x1=ppn\&(asid\textless{}\textless{}44)\&(mode\textless{}\textless{}60); \$csrw satp,x1;    \\ (2) FlushTLB: \$sfence.vma;\end{tabular}                                                                                         \\ \cline{2-3} 
			& SyncI            & \$fence.i;                                                                                                                                                                                \\ \cline{2-3} 
			& HFutex           & (Described in Section \ref{sec-hfutex})                                                                                                                                                               \\ \hline
			\multirow{3}{*}{\begin{tabular}[c]{@{}c@{}}Word\\ Data\\ Access\end{tabular}}  & RegRW            & (Using \emph{Reg} ports directly)                                                                                                                                                                          \\ \cline{2-3} 
			& MemR             & x1=addr; \$ld x2,x1(0); send x2;                                                                                                                                                         \\ \cline{2-3} 
			& MemW             & x1=addr; x2=data; \$sd x2,x1(0);                                                                                                                                                         \\ \hline
			\multirow{4}{*}{\begin{tabular}[c]{@{}c@{}}Page\\ Data\\ Access\end{tabular}}  & PageS            & x1=ppn\textless{}\textless{}12; x2=val; for (512) \{ \$sd x2,x1(0); \$addi x1,x1,8; \}                                                                                                     \\ \cline{2-3} 
			& PageCP           & \begin{tabular}[c]{@{}l@{}}x1=srcppn\textless{}\textless{}12; x2=dstppn\textless{}\textless{}12;\\ for (512) \{ \$ld x3,x1(0); \$sd x3,x2(0); \$addi x1,x1,8; \$addi x2,x2,8;\}\end{tabular} \\ \cline{2-3} 
			& PageR            & x1=ppn\textless{}\textless{}12; for (512) \{ \$ld x2,x1(0); \$addi x1,x1,8; send x2; \}                                                                                                    \\ \cline{2-3} 
			& PageW            & x1=ppn\textless{}\textless{}12; for (512) \{ recv x2; \$sd x2,x1(0); \$addi x1,x1,8; \}                                                                                                    \\ \hline
			\multirow{2}{*}{Perf}                                                          & Tick             & (Return overall ticks from system reset)                                                                                                                                                 \\ \cline{2-3} 
			& UTick            & (Return the total ticks in U-Mode for a given CPU from system reset)                                                                                                               \\ \hline
			\textit{Optional} & Interrupt            & (Using \emph{Interrupt} ports directly)                                                                                                               \\ \hline
		\end{tabular}
		\begin{tablenotes}
			\item[1] Relevant registers will be staged beforehand and restored upon completion via \emph{Reg} ports.
			\item[2] "=" means to set the value of a register via \emph{Reg} ports from \textbf{Arg Regs}.
			\item[3] "\$" means to inject an instruction via \emph{Inject} ports.
			\item[4] Controller blocks on the \textbf{Exception Event Queue} and dequeues CPU IDs to be processed. A CPU's ID will be queued when it switches from U- to M-Mode.
			\item[5] "Send" means to push a register value into \textbf{Resp Regs} (\textbf{TX Buffer} for \emph{PageR}) via \emph{Reg} ports.
		\end{tablenotes}
		\end{threeparttable}
	}
		\label{tab-htp}
	\end{table}

	
	HTP defines four types of host-initiated requests that cover the needs of remote system call handling:
	\textit{(1) Instruction Stream Management}
	Includes \textbf{Redirect}, \textbf{Next}, \textbf{MMU}, \textbf{SyncI}, and \textbf{HFutex}, which control instruction flow, exception retrieval, address translation state, synchronization, and hardware futex.
	\textit{(2) Word-Level Data Access}
	\textbf{RegRW} and \textbf{MemRW} provide machine-word-granularity access to CPU registers and physical memory.
	\textit{(3) Page-Level Data Access}
	\textbf{PageS} (PageSet), \textbf{PageCP} (PageCopy), and \textbf{PageRW} operate on full memory pages.
	\textit{(4) Performance Count}
	\textbf{Tick} and \textbf{UTick} return global and per-CPU execution ticks.
	
	Table \ref{tab-htp} summarizes all HTP requests and their handling methods through the predefined CPU interface.
	Bold entries (e.g., \textbf{Arg Regs}) map to hardware components in Figure \ref{fig-controller}, detailed in the next section.
	All HTP requests except \emph{Next} and \emph{Tick} specify a target CPU ID.
	Only CPUs stalled in privileged mode by \emph{StopFetch} may service HTP requests, preventing interference with user programs.
	All memory accesses are performed via injected load/store instructions and therefore comply with the target-hardware-defined memory model and coherence protocol.
	
	Experimental results show that HTP reduces UART traffic by over 95\% compared to direct CPU-interface calls.
	Although fine-grained requests (e.g., register access) incur modest header overhead, page-level operations, commonly used for page table transmission and copy-on-write handling, reduce communication volume to below 1\% of the direct approach.

\subsection{FASE Hardware Controller}
\label{section-controller-arch}

	\begin{figure} [t]
		\centering
		\includegraphics[]{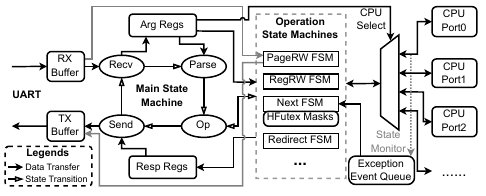}
		\caption{Architecture overview of FASE Hardware Controller.}
		\label{fig-controller}
	\end{figure}
	
	The FASE Controller is the central hardware component that bridges the interaction between host software and the FPGA-based target system.
	

	
	As shown in Figure \ref{fig-controller}, FASE controller adopts a two-level state machine design:
	(1) a main state machine for HTP request reception, transmission, and parsing, and (2) a set of operation-specific state machines processing individual request types according to Table \ref{tab-htp}.
	UART data are buffered to overlap communication with operation latency under back-to-back requests.
	\textbf{Arg Regs} and \textbf{Resp Regs} temporarily store the parameters and return values of the current request.
	The \emph{PageRW} state machine directly interacts with the UART buffer to stream page data.
	In addition, the \emph{Next} state machine incorporates hardware \emph{futex} handling logic, as described in Section \ref{sec-hfutex}, when a \emph{futex} system call is detected.
	
	For \emph{PageRW} operations, interleaving instruction injection with register accesses may introduce pipeline stalls due to register data availability.
	In practice, this overhead is amortized by batching operations, typically issuing consecutive accesses to 8 or 16 registers and injecting multiple load/store instructions within each iteration.

	The FASE Controller thus provides a robust and efficient interface for host-target communication, enabling syscall-level emulation with minimal performance and hardware cost.

\section{FASE Host Runtime and Syscall Handling}
\label{section-host}
	
	This section details the design of FASE host-side runtime, which is responsible for initializing target system and managing exceptions and system calls generated by the simulated application running on the target FPGA.
	As the core software component of the framework, this runtime provides the essential back-end logic and user interface for syscall emulation.
	
	The key challenge in this context lies in achieving sufficient system call coverage to support multi-thread/multi-process user-level workloads while relying solely on such HTP requests (described in Section \ref{section-controller-protocol}).
	The limited bandwidth and functional scope of HTP impose strict constraints on the design of the runtime.
	
	\begin{figure} [t]
		\centering
		\includegraphics[]{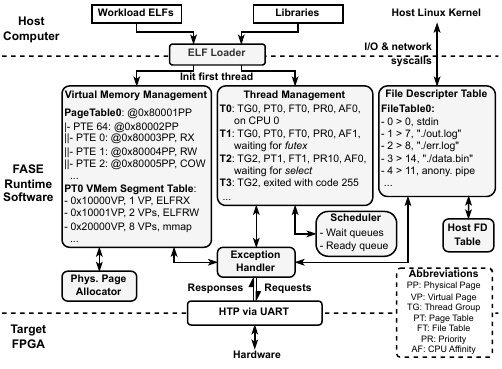}
		\caption{Architecture overview of FASE host-side runtime. An execution snapshot example is provided to illustrate the detailed data structures and correspondence among them.}
		\label{fig-software}
	\end{figure}

	\begin{figure*} [t]
		\centering
		\includegraphics[]{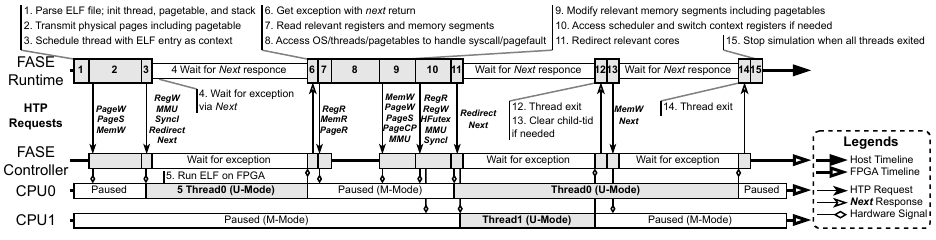}
		\caption{Full execution timeline of running a ELF file with 2 threads via FASE. Steps 6-11 show a typical workflow of remote system call processing using \emph{clone} as an example.}
		\label{fig-fulltimeline}
	\end{figure*}

	As illustrated in Figure \ref{fig-software}, the FASE host runtime consists of three key components, each addressing a fundamental requirement for syscall emulation: 
	thread scheduling and synchronization (Section \ref{section-rt-thread}), virtual memory management (Section \ref{section-rt-vm}), and I/O access bypass (Section \ref{section-rt-io}).
	As a interface component, the exception handler invokes three functional components to process syscalls and constructs associated HTP requests.

	Figure \ref{fig-fulltimeline} illustrates the execution timeline for user programs and syscall handling.
	After reset, all CPUs are in privileged mode and paused by the \emph{StopFetch} signal.
	Execution starts with a \emph{Redirect} to enter user mode, followed by \emph{Next} to wait for the next hardware exception.
	The FASE runtime maintains complete thread context, including the thread control blocks, page tables, virtual memory segments, and physical addresses of page table entries.
	Upon an exception, the runtime parses the CPU ID and exception metadata returned by \emph{Next}, identifies the thread, processes the syscall, and applies updates to page tables, memory, and registers to the target system.
	During user-mode execution on the FPGA, address translation and related mechanisms are handled entirely in hardware without host interaction.
	
	Not every syscall triggers a \emph{Redirect}.
	For instance, after an \emph{exit}, the thread is terminated; if no ready thread exists, no \emph{Redirect} is issued, and the CPU remains paused until a new thread is scheduled.

\subsection{Thread Scheduling and Synchronization}
\label{section-rt-thread}
	
	
	
	Upon a \emph{Redirect}, the CPU resumes execution solely from the supplied context, without awareness of thread identity.
	Scheduling a thread onto a paused CPU is therefore realized by storing current thread's context, loading a new ready thread’s context, and issuing a \emph{Redirect}.
	If ready threads outnumber paused CPUs, a subset is selected by the scheduling policy and the remainder stay in the ready queue.
	
	By default, FASE adopts a non-preemptive scheduling model sufficient for common computational workloads.
	A running CPU can only undergo a context switch after raising its next exception including timer interrupt.
	Optionally, the runtime accesses target hardware timers, if available, through specific MMIO addresses.
	Implement of \emph{interrupt} in FASE CPU interface would support more preemptive policies.

	In addition to conventional thread blocking and scheduling, two special mechanisms must be addressed: Linux signals and host-blocking system calls.
	
	Signals are asynchronous software interrupts used in Linux for inter-thread and inter-process communication.
	A thread may setup signal handlers for specific signal numbers. 
	When another thread sends a signal, the target thread's execution is redirected to the associated handler function if set.
	
	\begin{figure} [t]
		\centering
		\subfigure[Signal mechanism is triggered.]{
			\begin{minipage}{\columnwidth}
				\centering
				\includegraphics[]{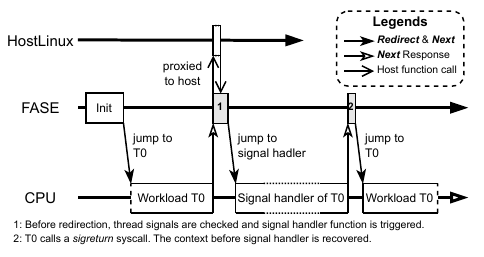}
			\end{minipage}
			\label{fig-syscall1-signal}
		}
		\subfigure[Some syscalls may block in the host kernel.]{
			\begin{minipage}{\columnwidth}
				\centering
				\includegraphics[]{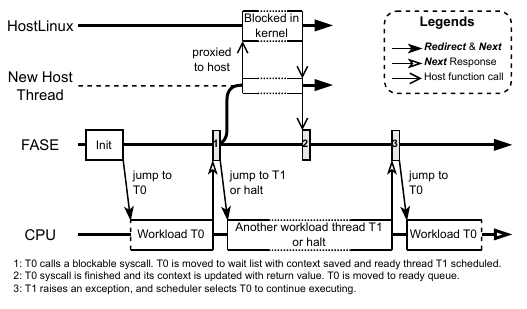}
			\end{minipage}
			\label{fig-syscall2-blocked}
		}
		\caption{Example of special scheduling situations.}
	\end{figure}

	As shown in Figure \ref{fig-syscall1-signal}, before resuming a thread, the scheduler checks the pending signal queue.
	If a signal is awaiting handling, the scheduler issues a \emph{Redirect} to a preloaded signal handling trampoline in target memory.
	This trampoline wraps the actual handler and invokes the \emph{sigreturn} system call upon completion to resume normal execution.
	
	Some I/O-related system calls are inherently block-able, like \emph{read}, \emph{select}, and \emph{recv}.
	When these calls are bypassed to the host, the FASE runtime itself will block in the host’s kernel space, halting the simulation.
	To mitigate this, the runtime will employ an auxiliary host thread to handle such blocking operations asynchronously, as depicted in Figure \ref{fig-syscall2-blocked}.
	This allows the runtime to remain responsive and continue managing other simulated threads without interruption.
	
	With these mechanisms in place, FASE supports most common Linux thread communication and synchronization primitives, to enable syscall emulation on multi-thread workloads.

\subsection{Hardware-Assisted Futex}
\label{sec-hfutex}

	The Linux kernel provides the \emph{Futex} (fast userspace mutex) mechanism as a low-overhead primitive for thread synchronization, decoupling user-space locking from kernel-side blocking.
	Threads use atomic operations to decide blocking conditions, invoking \emph{futex wait} on a memory address, while others issue \emph{futex wake} on same address to resume blocked threads.
	
	In a full Linux environment, \emph{futex wake} incurs low cost, involving only mode switches and a wait-table lookup.
	Consequently, thread libraries such as \emph{pthread} employ aggressive wake-up policies to wake up any threads that might blocked or to reset a \emph{futex}.
	In contrast, within FASE, each syscall requires UART transmissions, making redundant \emph{futex wake} operations highly expensive.
	
	To mitigate this, we introduce Hardware-Assisted Futex (HFutex), which allows the FASE controller to locally filter wake-ups.
	Each CPU core maintains a small \emph{HFutex Mask Cache}, updated by the host via dedicated HTP \emph{HFutex} requests.
	On the \emph{futex wake} address hit, the controller directly returns success code 0 and avoids unnecessary latency.

	\begin{figure} [t]
		\centering
		\includegraphics[]{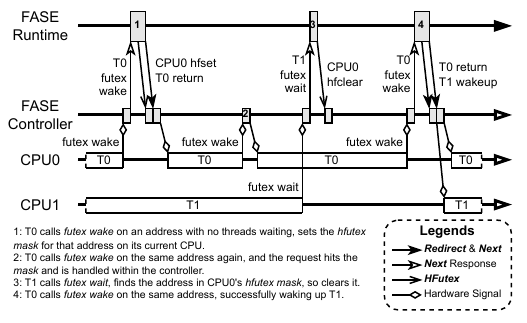}
		\caption{Example of HFutex operation across two threads on two CPU cores.}
		\label{fig-hfutex}
	\end{figure}
	
	Figure \ref{fig-hfutex} illustrates an example of HFutex operation. 
	When a \emph{futex wake} without waking any thread is handled by the runtime, its target address is added to the HFutex Mask of current core. 
	Meanwhile, both the physical and virtual target addresses are recorded in the runtime. 
	When a \emph{futex wait} succeeds, the HFutex Masks on all cores that contain the corresponding physical address are cleared. 
	Additionally, all HFutex Masks on a core are cleared upon a thread switch.

	By eliminating unnecessary system call round-trips for common futex patterns, HFutex significantly reduces the overhead on multi-thread syscall emulation. 
	
\subsection{Virtual Memory Management}
\label{section-rt-vm}
	
	The virtual memory subsystem manages page tables and memory-related system calls, ensuring that user programs on the target FPGA behave consistently with Linux semantics.
	
	Since FASE accesses target memory only through the low-bandwidth HTP interface, it employs additional mechanisms to minimize cross-device traffic, including dual software/hardware page tables, copy-on-write (COW), lazy initialization, and file preloading.
	
	
	

	A reference-counted page allocator manages device physical pages as the basic of memory management.
	Page tables are allocated on device memory, while the runtime maintains a complete software representation of each page-table.
	Virtual-memory-related system calls (e.g., \emph{mmap}, \emph{munmap}, and \emph{brk}) are processed based solely on the runtime metadata, with resulting updates to page tables and memory synchronized to the device via HTP.
	
	For page faults, copy-on-write (COW), and \emph{mmap} lazy initialization, the runtime also maintains a virtual segment table recording permission, associated file, file offset, and physical page mappings.
	Upon a page fault, the runtime locates corresponding segment by failed address, validates the access, allocates and initializes a physical page based on the address offset.
	
	For each file opened with \emph{mmap} (including anonymous shared memory treated as temporary files), sufficient physical pages are allocated and bound to the file.
	Shared mappings of the same file therefore access identical underlying contents.
	Frequently used files, such as dynamic libraries, may be pre-loaded and bound to physical pages to reduce \emph{mmap} overhead.
	
	As remote TLB invalidation based on inter-processor interrupt is optionally unsupported, remote TLB flush is delayed when target CPU traps into next exception, and runtime enforces non-overlapping virtual memory allocation to ensure correct execution of dangling-pointer-free workloads.
	
	Together, these mechanisms allow FASE to emulate a Linux-style virtual memory subsystem for multi-thread workloads.
	
\subsection{I/O Syscall Bypass}
\label{section-rt-io}

	The I/O syscall bypass component allows FASE to redirect I/O requests from the target FPGA to the host, eliminating the need for FPGA-based peripherals and simplifying hardware configuration.
	Target workloads thus interact directly with the host file system and devices.
	
	To maintain thread-level isolation, FASE runtime manages a file descriptor mapping table that links target-side descriptors to host files.
	Multiple threads may share the same table to enable inter-thread resource sharing.
	When a target I/O syscall is raised, the runtime translates the descriptor, invokes the corresponding host syscall, and returns the result via HTP.
	
	This mechanism provides a foundation for external workload interaction without hardware peripheral on the FPGA, and also supports advanced features such as \emph{mmap} and \emph{shm}.
	
	In summary, FASE integrates three core runtime components: thread scheduling and synchronization, virtual memory management, and I/O syscall bypass, along with auxiliary modules including the ELF loader, configuration database, and performance recorder.
	Together, they provide an full-stack framework for accurate performance evaluation on FPGA platforms.

\section{Experiments and Evaluation}
\label{section-exp}

	
	
	
	This section evaluates the accuracy and efficiency of \textbf{FASE} on the open-source RISC-V SMP Berkeley Rocket Core. 
	As baseline, we use a LiteX-based SoC integrating the same Rocket core\footnote{https://github.com/enjoy-digital/litex}\footnote{https://github.com/litex-hub/linux-on-litex-rocket}. 
	Both systems execute identical workload ELFs. 
	A single-core \textbf{Proxy Kernel (PK)} baseline is also implemented using Chipyard.
	
	The experiments pursue four objectives:
	\begin{itemize}
		\item[1.] Measure FASE’s accuracy against LiteX on benchmark scores and execution CPU time.
		\item[2.] Analyze per-benchmark errors and workload sensitivity.
		\item[3.] Evaluate the effects of UART baud rate and hardware-assisted futex.
		\item[4.] Assess early-stage validation efficiency via CoreMark and wall-clock comparisons.
	\end{itemize}
	
	All experiments use an open-source RISC-V processor for reproducibility. 
	Porting FASE to other ISAs requires adapting syscall register convention, page-table structure, and injected instruction accordingly.

\subsection{Experiment Setup}
	
\subsubsection{Target Processor Setup}
	
	\begin{figure} [t]
		\centering
		\includegraphics[width=\linewidth]{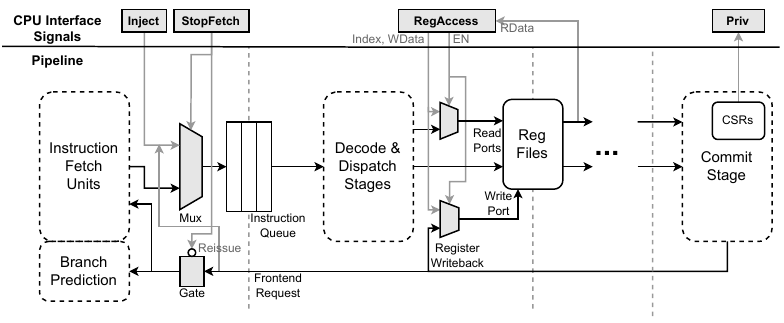}
		\caption{Modification on Rocket core pipeline to expose the FASE CPU interface. Shaded blocks and gray lines denote the added logic.}
		\label{fig-cpu-impl}
	\end{figure}
	

	
	Only the modifications in Figure \ref{fig-cpu-impl} are required to expose the FASE interface in Rocket.
	The shaded blocks and gray lines indicate the added logic:
	
	\begin{itemize}
		\item \emph{Priv} is extracted from the CSR unit.
		\item \emph{Reg} reuses existing register file read/write ports via multiplexing.
		\item \emph{Inject} is implemented with a clutch-like control on instruction queue (IQ); when \emph{StopFetch} is asserted, IQ is supplied only through \emph{InjectInst}.
		\item \emph{Inject ready} is asserted when the pipeline is empty, allowing single-instruction injection to comply with Rocket’s reissue behavior under stalls.
	\end{itemize}
	
	Single-instruction injection increases the FASE controller latency for HTP handling, but the overhead is negligible compared to UART transmission in the streaming protocol.
	
\subsubsection{Experimental Platform Setup}
	
	\begin{figure} [t]
		\centering
		\includegraphics[width=0.8\linewidth]{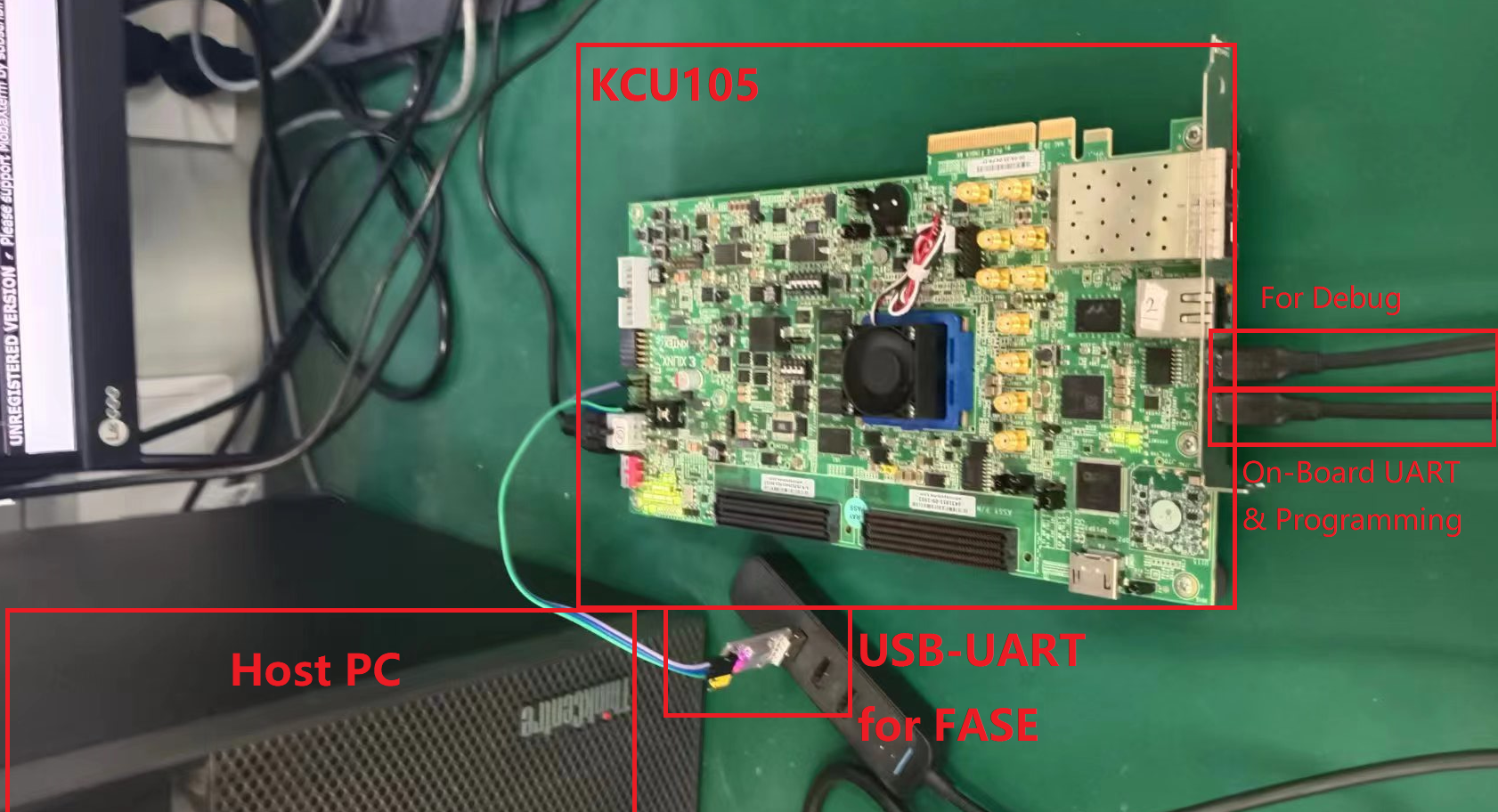}
		\caption{Hardware Devices Used in the Experiments.}
		\label{fig-exp-hw}
	\end{figure}

	
	The platform consists of a host PC and a Xilinx KCU105 FPGA (Figure \ref{fig-exp-hw}).  
	In \textbf{FASE}, an external USB–UART interface is deployed to bypass the 460Kbps limit of on-board UART.  
	In the \textbf{LiteX} baseline, the FPGA boots Linux (firmware + \emph{ramfs} with benchmarks) via on-board UART.
	
	\begin{figure} [t]
		\centering
		\subfigure[Baseline full-system SoC]{
			\includegraphics[width=0.5\linewidth]{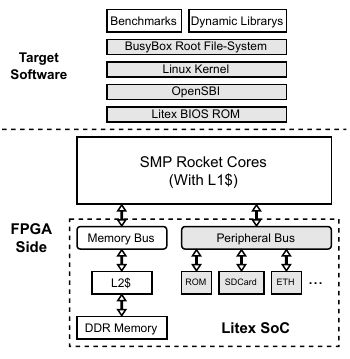}
			\label{fig-exparch-fs}
		}
		\subfigure[FASE without SoC]{
			\includegraphics[width=0.4\linewidth]{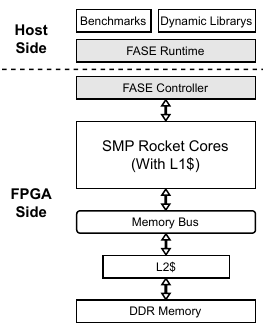}
			\label{fig-exparch-user}
		}
		\caption{Target hardware systems and software stacks for baseline Litex system and FASE system, respectively.}
		\label{fig-exparch}	
	\end{figure}


	\begin{table}[t]
		\centering
		\caption{Target Hardware and Host System Configurations}
		\resizebox{0.8\linewidth}{!}{
		\begin{tabular}{|c|c|l|}
			\hline
			\multirow{8}{*}{\textbf{\begin{tabular}[c]{@{}c@{}}Target\\ Hardware\end{tabular}}} & \textbf{Processor}      & Rocket Core v1.5, 1/4 SMP core(s)         \\ \cline{2-3} 
			& \textbf{Clock}          & 100MHz (125MHz for DDR)                   \\ \cline{2-3} 
			& \textbf{ISA}            & RV64GC, SV39 paged virtual memory         \\ \cline{2-3} 
			& \textbf{L1Cache}        & 32KB, 8-way for both L1I and L1D          \\ \cline{2-3} 
			& \textbf{Coherency}      & Tilelink coherent bus within Rocket       \\ \cline{2-3} 
			& \textbf{L2Cache}        & 256KB, 8-way, shared                      \\ \cline{2-3} 
			& \textbf{Memory}         & 2GB DDR4                                  \\ \cline{2-3} 
			& \textbf{FASE UART}      & 921600 bps baud-rate, 8N2 frame           \\ \hline
			\multirow{5}{*}{\textbf{\begin{tabular}[c]{@{}c@{}}Host\\ PC\end{tabular}}}         & \textbf{Processor}      & Intel(R) Core(TM) i9-14900K               \\ \cline{2-3} 
			& \textbf{OS}             & Ubuntu 22.04.4 with Linux 5.15.0          \\ \cline{2-3} 
			& \textbf{Memory}         & 128GB DDR5                                \\ \cline{2-3} 
			& \textbf{Host Compiler}  & gcc/g++ 13.2.0 (-O2 by default)           \\ \cline{2-3} 
			& \textbf{Cross Compiler} & riscv64-linux-gnu-gcc/g++/gfortran 13.2.0 \\ \hline
		\end{tabular}
		}
		\label{tab-conf}
	\end{table}


	Figure \ref{fig-exparch} summarizes the target system components.  
	The \textbf{LiteX} baseline implements a Linux-based full SoC.  
	\textbf{FASE} removes peripherals while retaining the memory subsystem (AXI, L2, DDR) and executes ELFs via the host runtime.
	Table \ref{tab-conf} lists detailed target hardware and host configurations.  
	LiteX peripherals use default settings, and other Rocket parameters follow the pre-generated LiteX CPU repository\footnote{https://github.com/litex-hub/pythondata-cpu-rocket}.
	
\subsubsection{Benchmarks}

	

	
	For single-core workloads, CoreMark is used.
	The \textbf{PK} baseline runs only on simulators, precluding complex benchmarks.
	
	For multi-core experiments on \textbf{FASE} and \textbf{LiteX}, we use the open-source GAPBS suite\footnote{https://github.com/sbeamer/gapbs}, which provides six OpenMP-based parallel graph algorithms: Betweenness Centrality (BC), Breadth-First Search (BFS), Connected Components via Shiloach–Vishkin (CCSV), PageRank (PR), Single-Source Shortest Paths (SSSP), and Triangle Counting (TC).
	
	GAPBS is cross-compiled on the host using \emph{riscv64-linux-gnu-g++} with dynamic linking and \emph{-O3}.
	The number of threads is set via \emph{OMP\_NUM\_THREADS} environment variable.
	Inputs are Kronecker graphs with $2^{20}$ vertices.
	Each run performs graph generation followed by 20 iterations, and the average per-iteration time is reported.

\subsection{Results on Performance Evaluation Accuracy}

	\begin{figure*} [!t]
		\centering
		\subfigure[GAPBS benchmark scores of FASE and LiteX baseline SoC across different thread numbers.]{
			\begin{minipage}{\linewidth}
				\centering
				\includegraphics[width=0.6\linewidth]{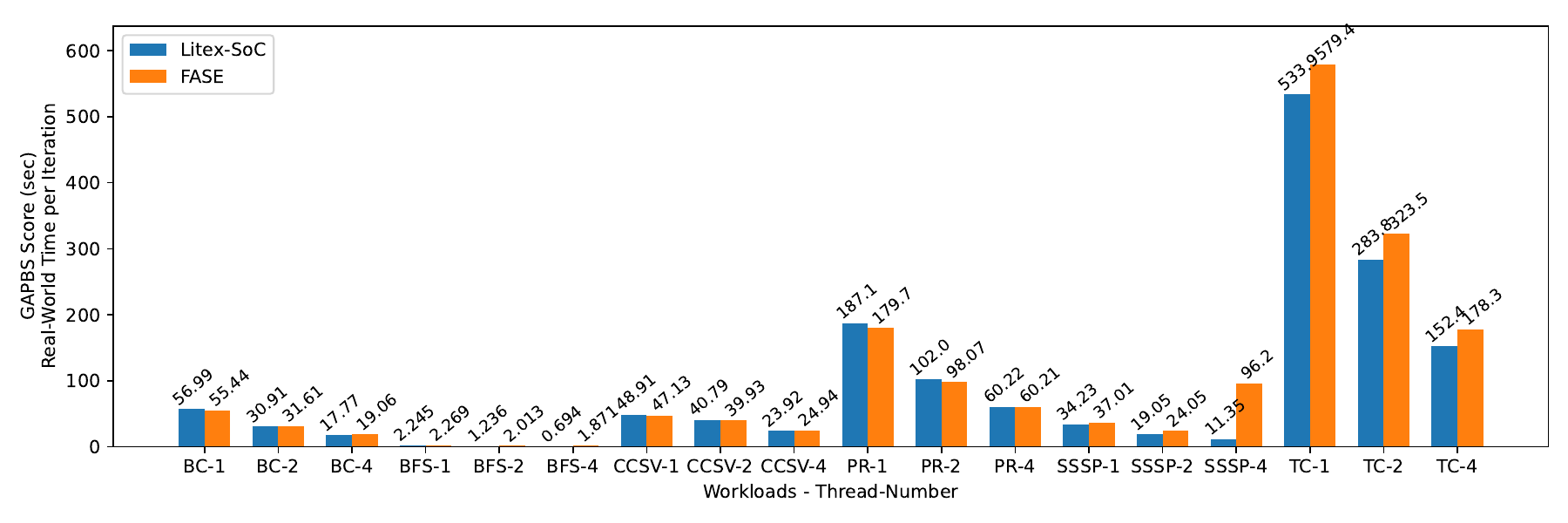}
			\end{minipage}
			\label{fig-data-com-score}
		}
		\subfigure[Total user CPU time records for each benchmark program of FASE and LiteX baseline SoC across different thread numbers.]{
			\begin{minipage}{\linewidth}
				\centering
				\includegraphics[width=0.6\linewidth]{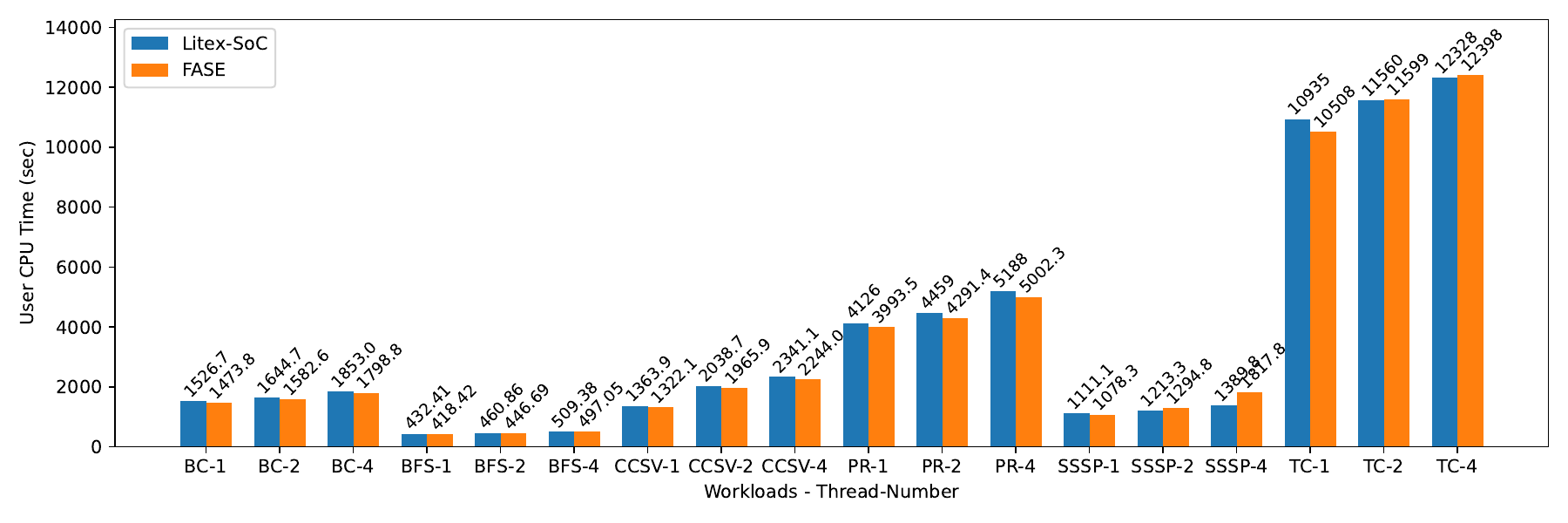}
			\end{minipage}
			\label{fig-data-com-user}
		}
		\subfigure[Relative error rates of GAPBS scores and user CPU time.]{
			\begin{minipage}{\linewidth}
				\centering
				\includegraphics[width=0.6\linewidth]{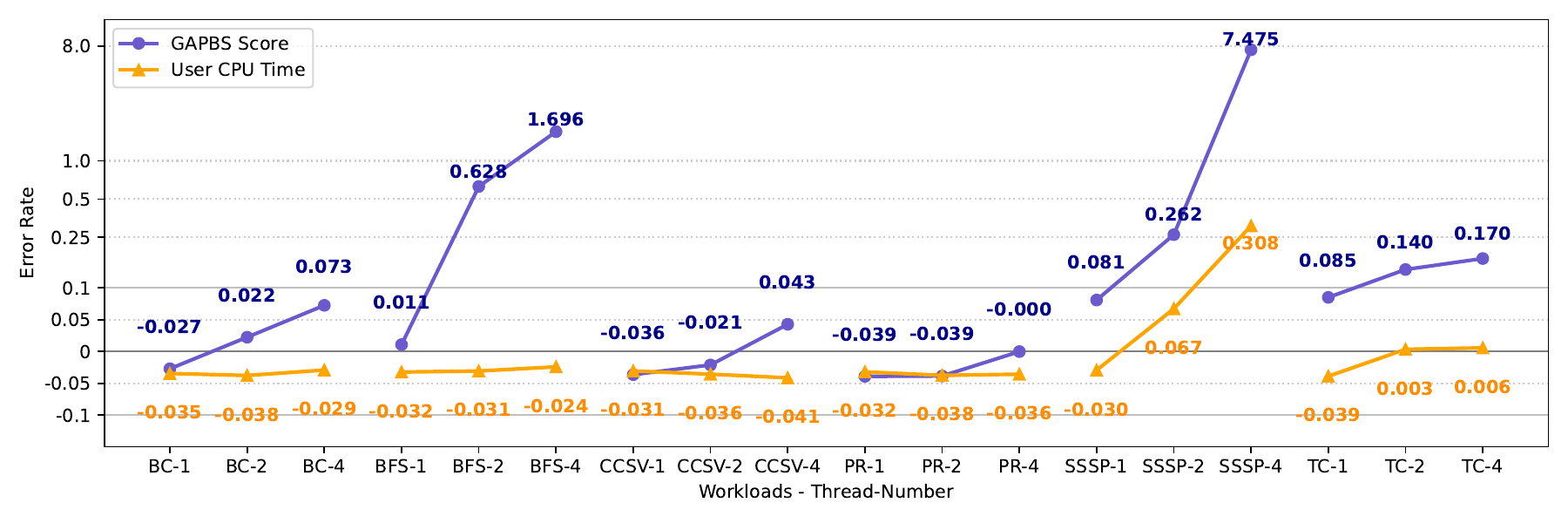}
			\end{minipage}
			\label{fig-data-com-err}
		}
		\caption{Comparative performance evaluation results of FASE and the LiteX baseline SoC in GAPBS benchmarks.}
		\label{fig-data-compare}	
	\end{figure*}

	\begin{figure*} [!t]
		\centering
		\subfigure[BC - Grouped by HTP requests.]{
			\begin{minipage}{0.23\linewidth}
				\centering
				\includegraphics[width=\linewidth]{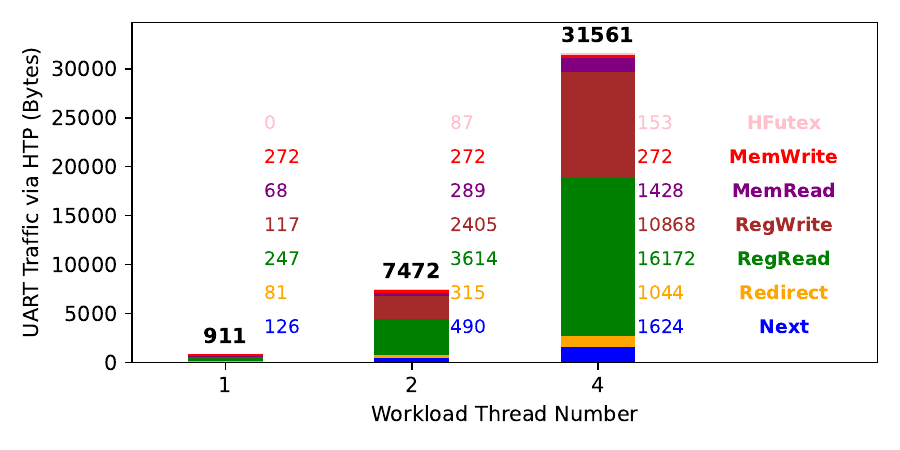}
			\end{minipage}
			\label{fig-data-bc1htp}
		}
		\subfigure[BC - Grouped by syscall types.]{
			\begin{minipage}{0.23\linewidth}
				\centering
			\includegraphics[width=\linewidth]{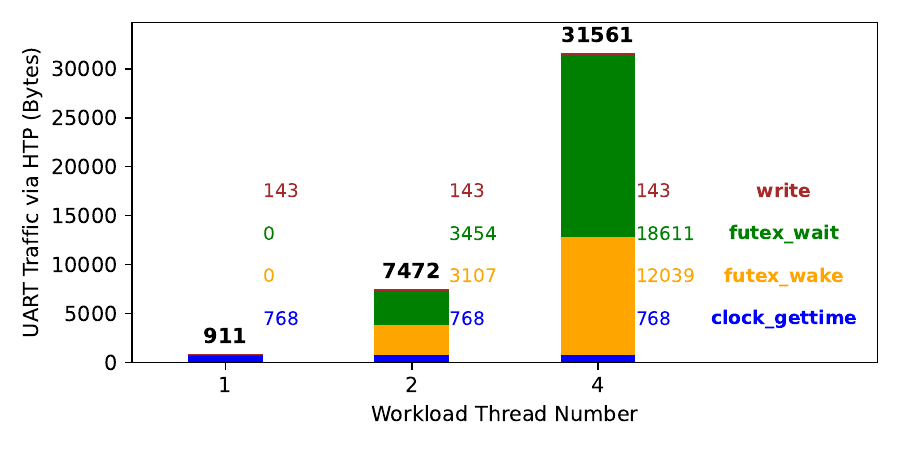}
		\end{minipage}
			\label{fig-data-bc1sys}
		}
		\subfigure[BFS - Grouped by HTP requests.]{
			\begin{minipage}{0.23\linewidth}
				\centering
			\includegraphics[width=\linewidth]{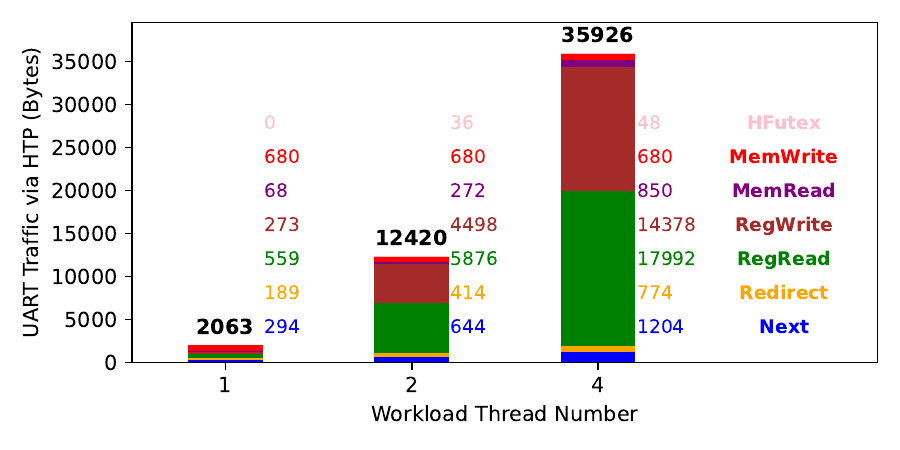}
		\end{minipage}
			\label{fig-data-bfs1htp}
		}
		\subfigure[BFS - Grouped by syscall types.]{
			\begin{minipage}{0.23\linewidth}
				\centering
			\includegraphics[width=\linewidth]{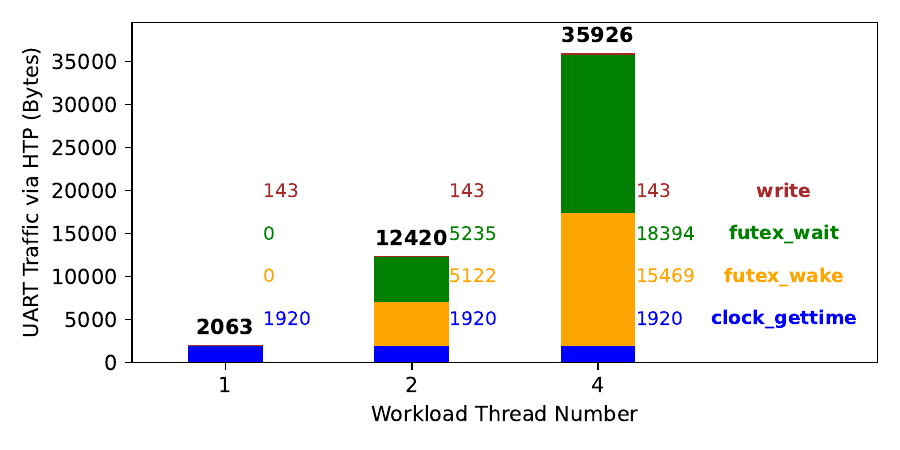}
		\end{minipage}
			\label{fig-data-bfs1sys}
		}
		\subfigure[SSSP - Grouped by HTP requests.]{
			\begin{minipage}{0.23\linewidth}
				\centering
			\includegraphics[width=\linewidth]{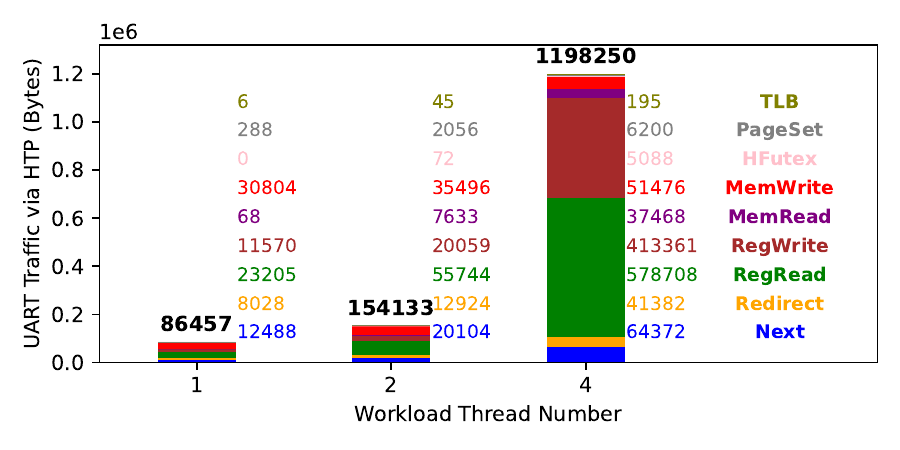}
		\end{minipage}
			\label{fig-data-sssp1htp}
		}
		\subfigure[SSSP - Grouped by syscall types.]{
			\begin{minipage}{0.23\linewidth}
				\centering
			\includegraphics[width=\linewidth]{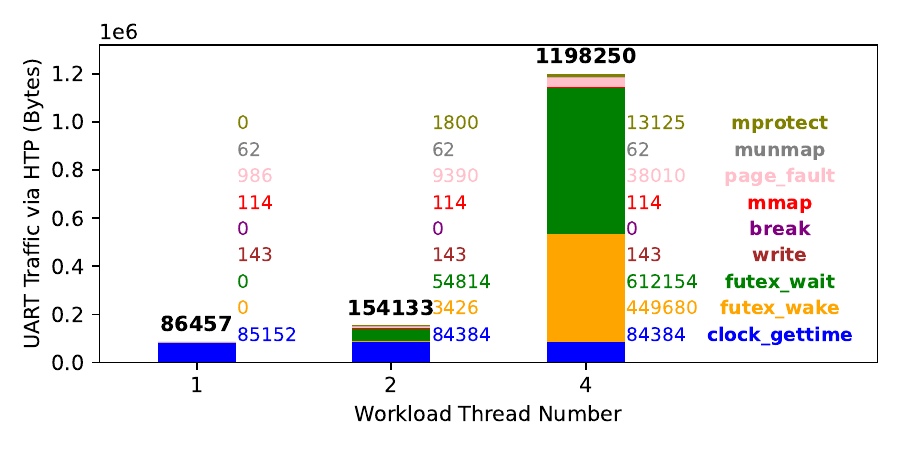}
		\end{minipage}
			\label{fig-data-sssp1sys}
		}
		\subfigure[TC - Grouped by HTP requests.]{
			\begin{minipage}{0.23\linewidth}
				\centering
			\includegraphics[width=\linewidth]{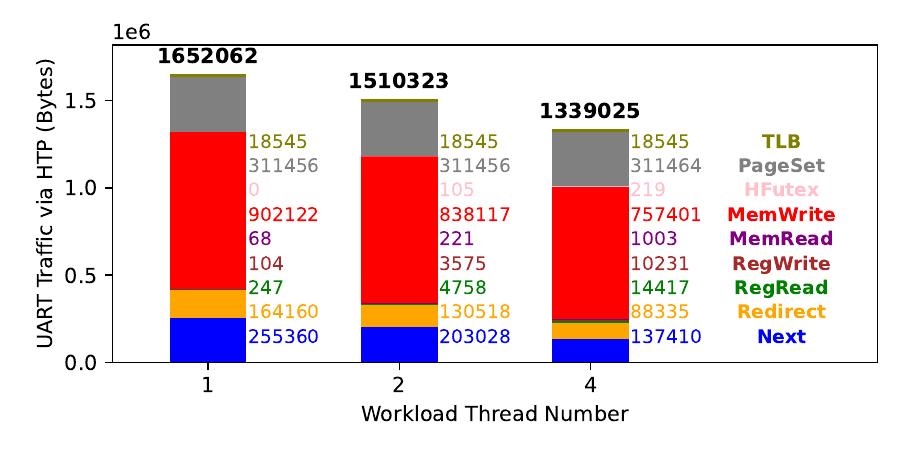}
		\end{minipage}
			\label{fig-data-tc1htp}
		}
		\subfigure[TC - Grouped by syscall types.]{
			\begin{minipage}{0.23\linewidth}
				\centering
			\includegraphics[width=\linewidth]{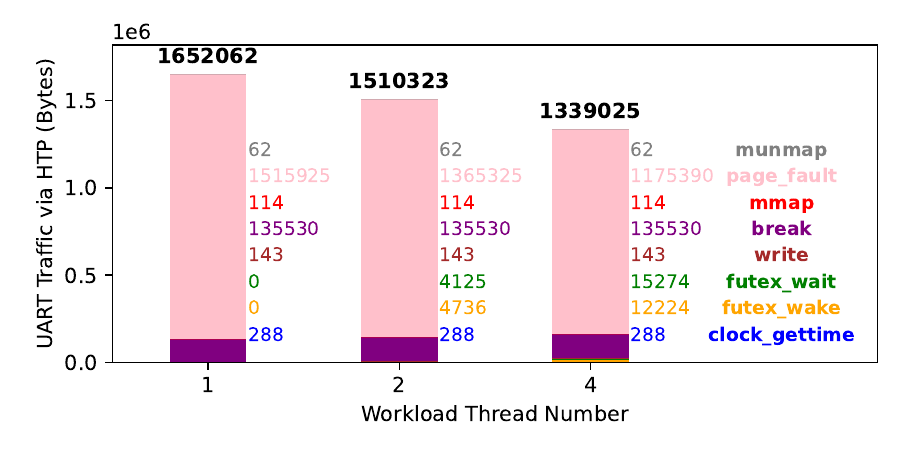}
		\end{minipage}
			\label{fig-data-tc1sys}
		}
		\caption{Composition of UART traffic for each workloads grouped by HTP requests and remote system call types. Tabular labels indicate the exact values of each component, with text colors matching the corresponding bar segments.}
		\label{fig-data-htp}	
	\end{figure*}

	This section presents and analyzes the performance differences between \textbf{FASE} and the \textbf{LiteX} baseline under identical processor hardware and benchmark workloads.
	For each benchmark and thread number, two metrics were recorded:
	\begin{itemize}
		\item[1.] \textbf{GAPBS score}: the average real-world time consumption per-iteration, measured within the benchmark program using the \emph{clock\_xxx} system calls and reported via \emph{stdout} upon completion.
		\item[2.] \textbf{User CPU time}: the total CPU time spent in user mode. In the \textbf{LiteX} baseline, this was obtained using the \emph{time} command; in \textbf{FASE} it was measured by the FASE controller.
	\end{itemize}
	For a data pair $(T_{fs}, T_{se})$, where $T_{se}$ and $T_{fs}$ denote the execution time reported by \textbf{FASE} and the \textbf{LiteX} full-system baseline, respectively, the relative error rate is defined as $e = (T_{se}-T_{fs})/T_{fs}$.

	Figures \ref{fig-data-com-score} and \ref{fig-data-com-user} show the GAPBS scores and user CPU times for all benchmarks and 1/2/4 thread numbers, while Figure \ref{fig-data-com-err} depicts the relative error rates.
	
	For single-thread workloads, FASE achieves low GAPBS score errors, below 3.9\% in four benchmarks and below 8.5\% in the remaining two.
	Error increases with thread number due to more frequent system calls from multi-threaded synchronization.
	The magnitude of this growth varies by benchmark.
	BC, CCSV, PR, and TC show moderate increases ($<$10\%), while BFS and SSSP exhibit unacceptable errors at four threads.
	This stems from benchmark-specific computation and synchronization patterns, analyzed in the following section.
	
	For most workloads, the user CPU time error remains near –3\%, indicating slightly reduced user-mode execution in FASE.
	This arises because (1) benchmark processes are isolated from other applications or kernel activities, reducing transient resource invalidations (e.g., TLB and cache), and (2) kernel-level time accounting introduces small delays before returning to user mode.
	In SSSP and TC, error grows with thread number due to syscall latency that extends user-mode synchronization across threads.
	
	Overall, FASE achieves over 91.5\% accuracy for GAPBS scores and over 95.9\% for user CPU time across most workloads.
	The following analysis examines how suboptimal parallel patterns in BFS, SSSP, and TC produce errors that are particularly challenging for FASE.

\subsection{Error Composition Analysis}

	Building on the previous section, we focus on BFS, SSSP, and TC, the workloads with the highest error rates under multi-threaded execution, and conduct additional experiments to validate the findings.

	Directly measuring remote system call or HTP request latency would introduce significant perturbations, so we use UART bytes transmitted as an indicator for error composition.
	In FASE, the dominant overhead comes from UART transfers.
	For instance, at 1 Mbps with 8N2 framing, transmitting a 40-byte physical page number and 64 bytes of data requires 1.144 ms, whereas the controller completes a \emph{PageSet} operation via the \emph{Inject} interface in 0.01 ms at 100 MHz.
	The host Linux kernel introduces only microsecond-scale delays to access the serial device buffer.

	Figure \ref{fig-data-htp} shows UART traffic per iteration for BC, BFS, SSSP, and TC, grouped by HTP requests and remote system call types.
	Data were sampled from the 10th iteration in twenty for each workload.
	BC serves as a control case, while the other three workloads highlight distinct factors contributing to FASE’s error.
	The following subsections provide a detailed breakdown.

	\begin{figure} [t]
		\centering
		\includegraphics[width=0.8\linewidth]{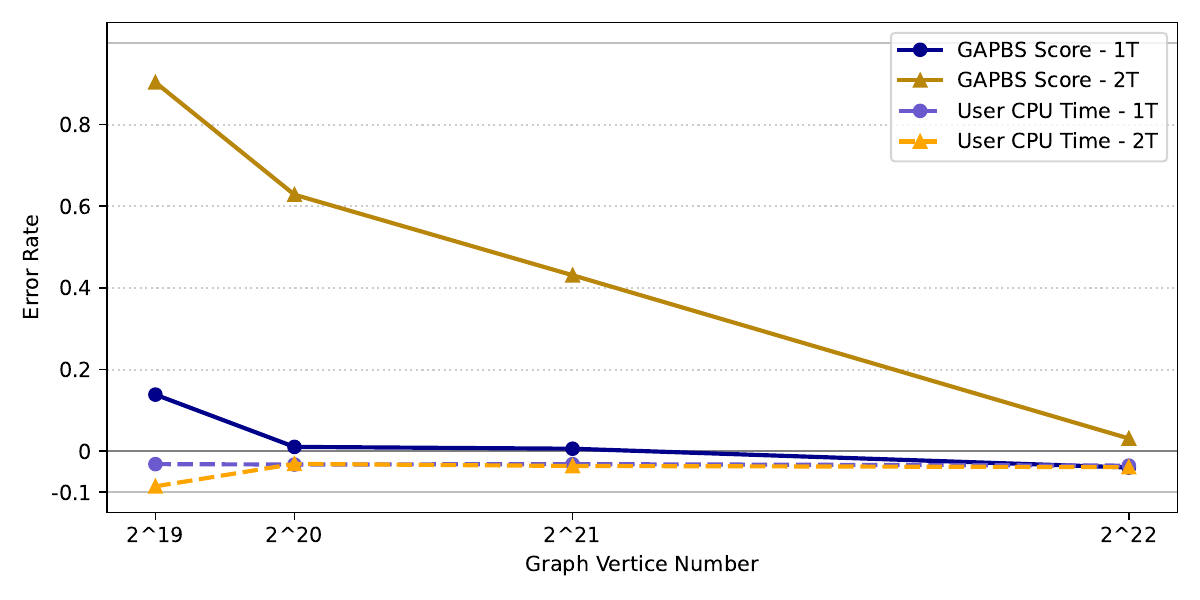}
		\caption{The error rate on BFS across different data scales. "GAPBS Score - 1T" denotes the error rate of GAPBS Score with 1 OpenMP thread.}
		\label{fig-data-bfsparam}
	\end{figure}

	\begin{figure} [t]
		\centering
		\includegraphics[width=0.8\linewidth]{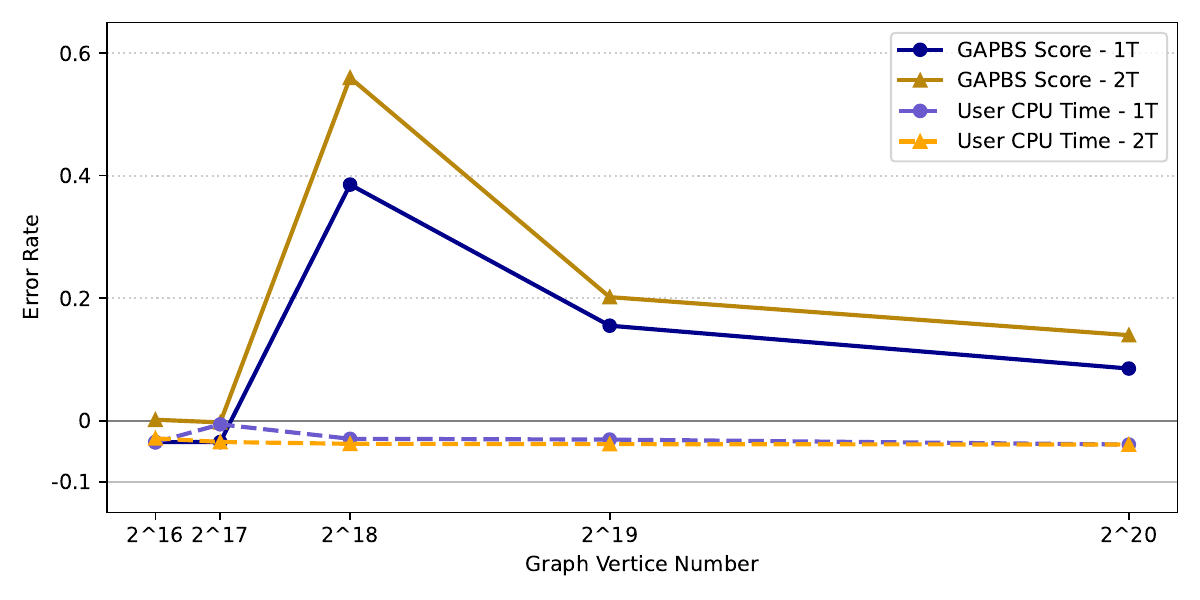}
		\caption{The error rate on TC across different data scales. "GAPBS Score - 1T" denotes the error rate of GAPBS Score with 1 OpenMP thread.}
		\label{fig-data-tcparam}
	\end{figure}

\subsubsection{For BFS: Minimal Computation Time Amplifies the Fixed Overhead}

	BFS has the smallest computational workload among the six benchmarks, with execution times only 1/10–1/100 of others.
	Although BC and BFS show comparable system call and HTP data volumes (Figures \ref{fig-data-bc1htp}, \ref{fig-data-bc1sys}, \ref{fig-data-bfs1htp}, \ref{fig-data-bfs1sys}) and similar absolute error magnitudes (Figure \ref{fig-data-com-score}), the user CPU time error rate of BFS-4 is 16.8× that of BC-4.
	Since multi-threading partially hides remote system call latency, this ratio consistent with the 20× difference in execution time between BC-4 and BFS-4.

	Figure \ref{fig-data-bfsparam} further confirms this trend across BFS data sizes.
	As the graph scale increases, per-iteration error decreases sharply.
	For $2^{22}$ nodes, 1/2-thread executions take 15.5 s and 8.9 s per iteration, with error dropping below 5\%.
	Similar to Linux, parallel performance degrades sharply when synchronization intervals approach synchronization overhead.
	However, FASE incurs additional overhead from remote system call latency, making it more sensitive to frequent synchronization.

\subsubsection{For SSSP: High-Frequency System Calls Destroy the User-Mode Synchronization}
\label{sec-exp-sssp}

	In SSSP, small synchronization blocks execute across many iterations, and each block’s run-time is measured individually in benchmark code, generating 40–400× more \emph{clock\_gettime} system calls (Figure \ref{fig-data-sssp1sys}).

	Most additional delay in SSSP-4 comes from synchronization overhead.
	Typically, threads first spin using user-mode atomic instructions and fall back to \emph{futex} only when spin-sync timeout.
	Remote system call latency delays spin completion, causing excessive \emph{futex} usage.
	This is evident in Figure \ref{fig-data-sssp1htp} and \ref{fig-data-sssp1sys}, since the context-switch overhead (reading/writing 63 registers) is 10–16× greater than the overhead of handling a \emph{futex} call itself (accessing only 4–7 argument registers).
	
	Baud-rate experiments also confirm this: reducing UART speed beyond a threshold sharply increases \emph{futex} overhead in SSSP-2, as \emph{clock\_gettime} latency exceeds spin-sync limits.

\subsubsection{For TC: Repeated Large-Scale Memory Allocations Introduce Initialization Overhead}

	In TC, each iteration allocates and releases large workspaces (128MB via \emph{mmap} and 4MB via \emph{break}), unlike other benchmarks that allocate once and reuse memory.
	Page faults from \emph{mmap} lazy initialization account for ~90\% of the cost, with \emph{break} initialization covering most of the remainder (Figure \ref{fig-data-tc1sys}).
	Hardware page table synchronization using \emph{MemWrite} contributes ~60\% of overhead, zeroing new pages using \emph{PageSet} takes ~25\%, and remaining page fault handling using \emph{Next}/\emph{Redirect} takes most of the rest (Figure \ref{fig-data-tc1htp}).

	FASE reduces \emph{Next}/\emph{Redirect} cost by preloading 16 pages per page fault, but total allocation remains large, leading to a 17\% error rate in TC-4.
	Error grows with data size, spiking at $2^{18}$ due to \emph{glibc} switching from heap-allocation to \emph{mmap} for large memory blocks (Figure \ref{fig-data-tcparam}), since reallocating heap is fully in user-space.
	
	Unlike SSSP, TC’s user CPU time estimation is less affected because lazy loading occurs during compute phases, evenly across threads.
	In SSSP, \emph{clock\_gettime} occurs near synchronization points, amplifying remote syscall latency.

\subsection{Further Validation on Configurable Factors}

	\begin{figure} [t]
		\centering
		\includegraphics[width=0.8\linewidth]{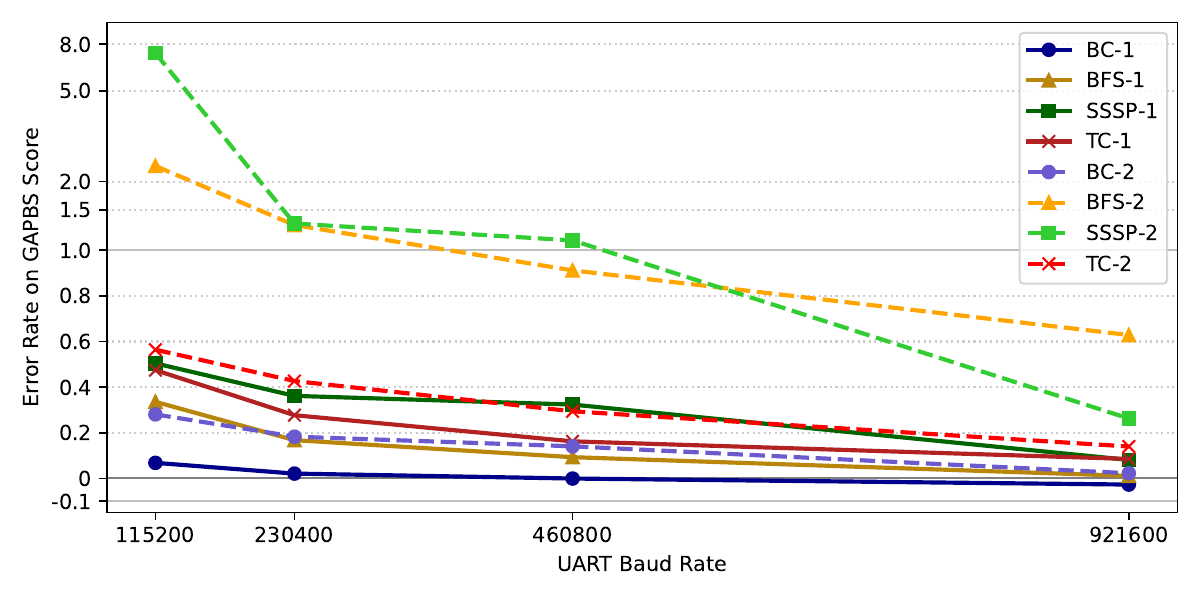}
		\caption{GAPBS score error rates on different UART baud-rate. "BC-1" means BC benchmark with 1 OpenMP thread.}
		\label{fig-data-baudrate}
	\end{figure}
	
	\begin{table}[t]
	\centering
	\caption{Breakdown of Stall Time per Iteration.}
	\resizebox{\linewidth}{!}{
		\begin{threeparttable}
		\begin{tabular}{|c|ccc|c|c|}
			\hline
			\multirow{2}{*}{\textbf{Workload}} & \multicolumn{3}{c|}{\textbf{Stall Time Composition (921600 bps)}}                                             & \multirow{2}{*}{\textbf{\begin{tabular}[c]{@{}c@{}}Theoretical\\ Stall Time\textsuperscript{2}\end{tabular}}} & \multirow{2}{*}{\textbf{\begin{tabular}[c]{@{}c@{}}Controller Stall Time\\ (Error rate) in Sim\textsuperscript{3}\end{tabular}}} \\ \cline{2-4}
			& \multicolumn{1}{c|}{\textbf{Controller}} & \multicolumn{1}{c|}{\textbf{UART}} & \textbf{Runtime\textsuperscript{1}} &                                                                                      & \\ \hline
			BC-1                               & \multicolumn{1}{c|}{7.98us}            & \multicolumn{1}{c|}{17.92ms}      & $\sim$183.07ms   & $\sim$183.08ms                                                                             & 7.98us(-0.029)                                                                                                  \\ \hline
			BC-2                               & \multicolumn{1}{c|}{42.13us}            & \multicolumn{1}{c|}{59.25ms}     & $\sim$828.48ms  & $\sim$828.52ms                                                                            & 32.67us(-0.032)                                                                                                  \\ \hline
			BC-4                               & \multicolumn{1}{c|}{166.03us}           & \multicolumn{1}{c|}{249.05ms}     & $\sim$1566.68ms  & $\sim$1566.84ms                                                                            & 61.14us(-0.044)                                                                                                 \\ \hline
		\end{tabular}
		\begin{tablenotes}
			\item[1] Runtime user-mode time $+$ average UART access latency $\times$ the number of UART accesses, to approximate the interval from each UART read to the subsequent write.
			\item[2] Sum of stall time introduced by the controller and runtime when transmission bandwidth is infinite.
			\item[3] Controller-induced stall time measured in cycle-accurate simulation, assuming instantaneous transmission and system call handling in hardware time.
		\end{tablenotes}
		\end{threeparttable}
	}
	\label{tab-stall-time}
	\end{table}
	
	\begin{figure*} [!t]
		\centering
		\includegraphics[width=0.7\linewidth]{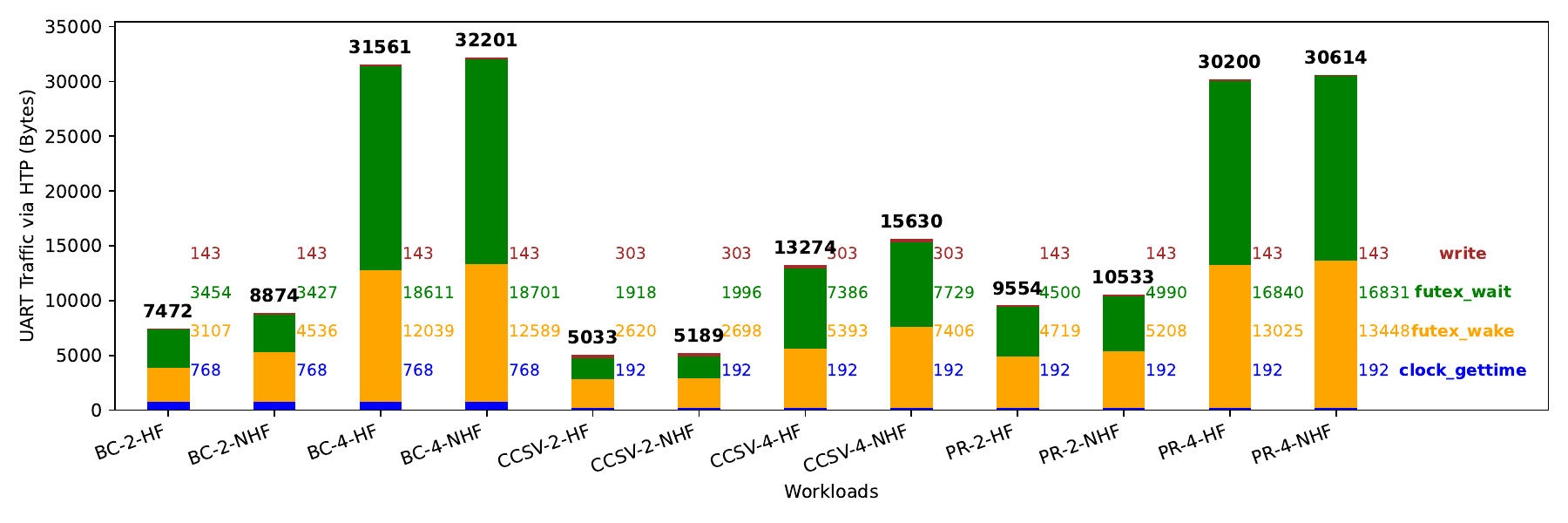}
		\caption{Impact of HFutex on UART traffic, grouped by remote system call type. “NHF” denotes HFutex NOT enabled, while “HF” denotes HFutex enabled.}
		\label{fig-data-hfutex}
	\end{figure*}

	In this section, the impact of two external factors on FASE’s performance accuracy is experimentally evaluated: UART baud-rate and hardware-assisted futex.

\subsubsection{Impact of UART Baud-Rate on Error Rate}

	Additional experiments were conducted to evaluate the effect of UART baud rate on error rate.
	For BC and the three benchmarks with the highest error rates, FASE’s GAPBS score error rates were measured under various baud-rates, as shown in Figure \ref{fig-data-baudrate}.
	
	For most benchmarks, the error rate decreases approximately linearly in a diminishing rate while baud-rate increasing.
	This aligns with the analysis that transmission overhead is a significant error component, while the residual error is attributed to the inherent overhead of remote system call handling and serial device access.
	Notably, SSSP-2 exhibited a severe error at 115200 bps, whereas a similar issue appeared only for SSSP-4 at higher baud-rates, corroborating and confirming the error source analysis of SSSP-4 in Section \ref{sec-exp-sssp}.

	Table \ref{tab-stall-time} presents the breakdown of stall time incurred by remote system calls under BC workload at 921600 bps.
	At this baud rate, transmission latency accounts for approximately 25\%.
	The dominant overhead, however, originates from the runtime, primarily due to host-side system calls triggered by UART accesses and file operations.
	Reducing UART access frequency by merging HTP requests and increasing transmission bandwidth are therefore critical to improving FASE accuracy in the future.
	
	To evaluate the ideal case, we further conducted simulation experiments where the HTP transmission and FASE runtime does not advance simulated time.
	That is, successive HTP requests become effective immediately upon completion in hardware time.
	Under the same four-thread workload, controller-induced stall time decreases by approximately 60\%, corresponding to fewer \emph{futex} invocations.
	This result supports the earlier analysis: delays in thread timelines cause some user space synchronize scenarios to exceed their timing window, thereby triggering \emph{futex}-based waiting.
	

\subsubsection{Verification of Hardware-Assisted Futex}

	Further experiments were conducted to evaluate the practical effectiveness of the \emph{HFutex} mechanism.
	To minimize the influence of other factors, the tests were performed on BC, CCSV, and PR benchmarks, which exhibit low error rates and involve only three system calls, \emph{futex}, \emph{write}, and \emph{clock\_gettime}.
	As in previous evaluations, the UART traffic was used as the primary metric for measuring optimization effectiveness.
	
	As shown in Figure \ref{fig-data-hfutex}, \emph{HFutex} influenced the \emph{futex}-related traffic as expected, primarily reducing the volume of \emph{futex\_wake} transmissions.
	The reduction in total traffic ranged from 3\% to 15\%, with the specific amount varying across programs.
	For instance, in BC-2, approximately 30\% of \emph{futex\_wake} calls were suppressed, whereas in CCSV-2, the improvement was negligible.
	This variation is attributed to differences in the parallel algorithms and the implementation details of the OpenMP library.

\subsection{Comparative Evaluation of FASE and Proxy Kernel on CoreMark}

	\begin{figure} [t]
		\centering
		\subfigure[Rocket.]{
			\begin{minipage}{0.62\columnwidth}
				\centering
				\includegraphics[trim=5pt 0 7pt 0, clip, width=\linewidth]{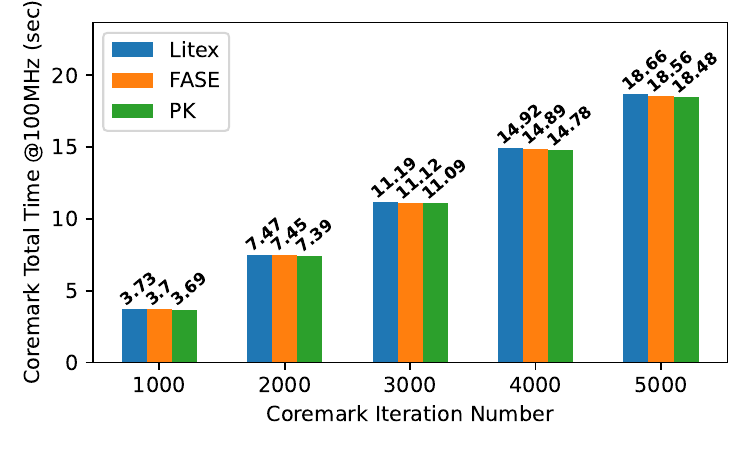}
			\end{minipage}
			\label{fig-coremark-score-rocket}
		}
		\subfigure[CVA6 (Single-Issue).]{
			\begin{minipage}{0.31\columnwidth}
				\centering
				\includegraphics[trim=5pt 0 5pt 0, clip, width=\linewidth]{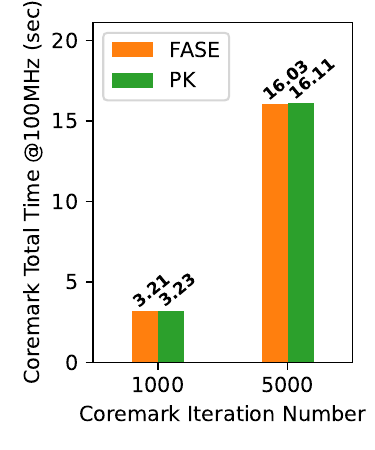}
			\end{minipage}
			\label{fig-coremark-score-cva6}
		}
		\caption{CoreMark-based single-core performance evaluation result with FASE, Litex, and PK (on Verilator). Execution time is recorded and reported by CoreMark program itself.}
		\label{fig-coremark-score}
	\end{figure}
	
	\begin{figure} [t]
		\centering
		\subfigure[PK baseline, where “2 Thread” indicates 2 Verilator simulation threads.]{
			\begin{minipage}{\columnwidth}
				\centering
				\includegraphics[width=0.7\linewidth]{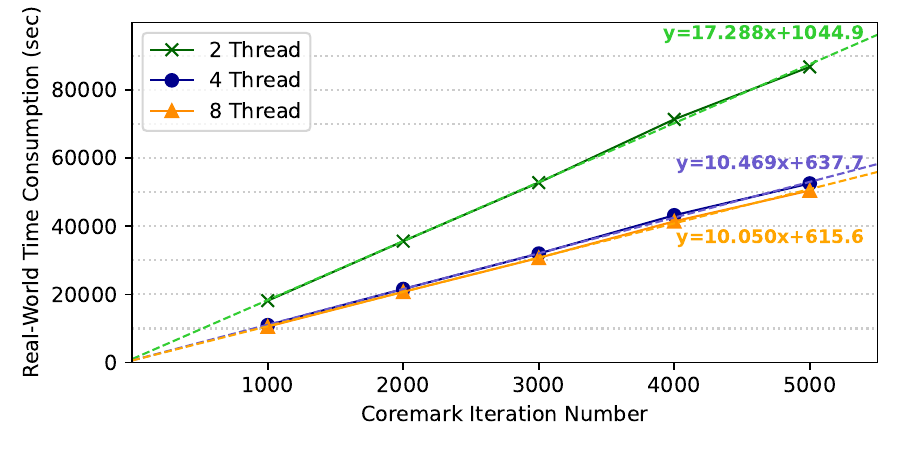}
			\end{minipage}
			\label{fig-coremark-time-pk}
		}
		\subfigure[FASE, where “115200 bps” indicates a UART baud rate of 115200.]{
			\begin{minipage}{\columnwidth}
				\centering
				\includegraphics[width=\linewidth]{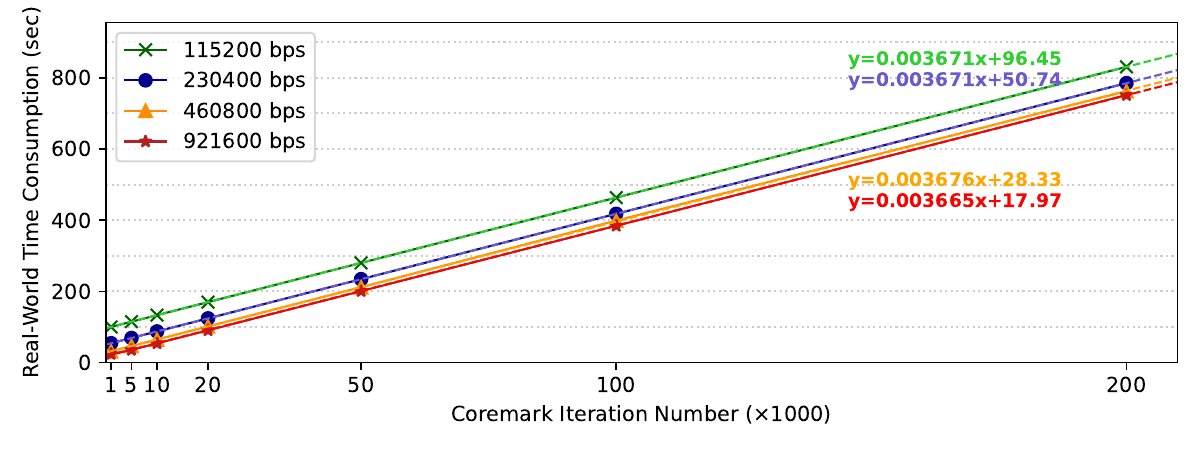}
			\end{minipage}
			\label{fig-coremark-time-fase}
		}
		\caption{Comparison of wall-clock time, i.e. real-world time, consumed by PK and FASE for executing CoreMark. Time records include system boot, workload loading, and workload executing, but excludes compilation before-hand.}
		\label{fig-coremark-time}
	\end{figure}
	
	In this section, the CoreMark results obtained by \textbf{FASE}, \textbf{Litex}, and \textbf{PK} are compared.
	In early-stage processor validation scenarios, FASE achieves higher performance accuracy and over a 2000x improvement in evaluation efficiency due to FPGA acceleration enabled.
	
	Figure \ref{fig-coremark-score-rocket} presents the performance results of the three systems running the same CoreMark program.
	Using the \textbf{Litex} full-system validation as the reference, \textbf{FASE} exhibits an error rate below 1\%, while \textbf{PK}’s error rate is approximately twice that of \textbf{Litex}.
	This is because \textbf{FASE} uses the same DDR hardware as the \textbf{Litex} full-system, whereas \textbf{PK} relies on simulated DDR components, whose detailed timing differs from the DDR on FPGA.
	We also evaluated FASE on a single-core CVA6 to validate its cross-microarchitecture generality as shown in Figure \ref{fig-coremark-score-cva6}, observing an error rate below 1\%.
	
	Figure \ref{fig-coremark-time} illustrates the runtime efficiency of \textbf{FASE} and \textbf{PK}.
	In the \textbf{PK} baseline using 8 Verilator threads, a single CoreMark iteration takes approximately 10 seconds of real-world time, whereas \textbf{FASE} completes the same iteration in only 0.0037 seconds, benefiting from substantial FPGA acceleration.
	In Figure \ref{fig-coremark-time-pk}, the intercept (i.e., startup time) scales roughly with simulator efficiency, since the startup overhead stems from PK executing initialization code on the simulated CPU.
	Workload loading in PK involves only host-side file access and incurs negligible time.
	Using 8 threads instead of 4 does not yield significant performance improvement, limited by the inherent parallelism of Verilator.
	In Figure \ref{fig-coremark-time-fase}, the intercept does not scale directly with UART baud rate, because FASE performs additional environment setup and data transferring verification during workload loading, leaving the UART stream underutilized.
	
	Overall, both FASE and PK target the same scenario: enabling early-stage processor validation with relatively complex non-baremetal workloads.
	PK offers the advantage of easy debugging based on RTL simulators.
	In contrast, FASE emphasizes much higher validation efficiency (over 2000× acceleration) and broader workload support (multi-threading, dynamic linking, and more complex system calls), while some debugging tasks are also supported by cooperating with FPGA integrated logic analyzers (ILA).

\section{Discussion and Limitation}
\label{section-discussion}

	We first assess whether FASE achieves its intended benefits: (1) early, accurate, and efficient performance evaluation, and (2) reduced engineering effort during agile processor design iteration.

	For the first point, although FASE runs on a full Rocket core pipeline in the experiments, only a minimal subset of hardware is actually exercised.
	At the instruction level, only user-mode instructions, \emph{sfence.vma} for TLB flushes, and \emph{mret} for exception return are used.
	At the CSR level, only \emph{satp} is used for page table setup, and \emph{mstatus}, \emph{mcause}, and \emph{mepc} for exception information.
	Providing additional optional CPU interface signals for exception status or page table configuration could completely eliminate the use of non-user-mode instructions.
	At the data level, only the main memory bus is utilized; uncached ports and MMIO are unused.
	This demonstrates that FASE can provide accurate performance evaluation even on minimal early-stage processor prototypes with only a minimal pipeline and a simplest LSU.
	Moreover, FASE’s functionality is unaffected by other hardware structures such as multi-core interconnects, further highlighting its broad applicability across microarchitectures.
	
	For the second point, in the experiments, only L2 cache and DDR are setup outside the core, without any peripheral integration.
	Cache and DDR components are typically supported natively by FPGA toolchains, making integration and debugging much simpler than peripheral IPs.
	Based on ASIC experience \cite{ref-asic-prototype, ref-cad-assisted-sim2, ref-hlpow}, such early validation can prevent about 65\% of late-stage design rework, with only ~10\% additional early-stage effort for FPGA adaptation.
	herefore, FASE effectively supports fast iteration process in agile processor design.

	We next discuss the remaining challenges and limitations of FASE.
	
	The first limitation stems from syscall emulation’s nature: it cannot model system-mode performance.
	Workloads with frequent syscalls and thread synchronization, such as SSSP and BFS in the experiments, exhibit significant accuracy loss.
	This issue is less pronounced in software-based syscall emulation, where syscalls complete nearly instantaneously for target system.
	But in FASE, remote syscall handling introduces additional latency.
	This can be mitigated by stalling cores during syscall processing or increasing cross-device communication bandwidth.
	

	The second limitation stems from the bandwidth of physical HTP channel.
	While a 921 Kbps UART is sufficient to maintain accuracy in a four-core system, communication demand increases with core count, and beyond roughly 12 cores, the bandwidth will become a dominant error source.
	Moreover, the centralized hardware controller may limit timing behavior on FPGA routing.
	Future work will explore higher-bandwidth interconnects (e.g., PCIe) and scalable designs with multiple controllers , multiple HTP channels, and multithread runtime to better support many-core systems.

\section{Related Works}
\label{section-relatedworks}

	In this section, we discuss and compare existing related work from three perspectives:
	(1) software-based syscall emulation methods;
	(2) early-stage processor performance evaluation techniques;
	and (3) end-to-end hardware evaluation approaches and workflow simplification.

\subsection{Software-Based Syscall Emulation}

	Syscall emulation (SE) was introduced in agile microarchitecture simulators to execute user programs while bypassing devices and target OS\cite{ref-survey-cachesim, ref-manycore-framework}.
	Prior work has validated its accuracy for low-I/O workloads when OS behavior is not the focus \cite{ref-esesc, ref-vmsim, ref-gem5}, typically for many-core system \cite{ref-zsim, ref-sniper}.
	McSimA+ \cite{ref-mcsima} speedups by executing the workload functionally and replaying instruction traces in a hardware timing model, whereas FASE executes user programs natively on target CPU for detailed and accurate performance insights.
	These tools support early-exploration in agile design but require extra software modeling effort and lack RTL-fidelity performance insight.
	
	In contrast, RTL-based simulators typically operate in full-system mode due to the absence of direct syscall interfaces \cite{ref-hwsw-rtl-sim, ref-fireaxe-large-rtl-sim}.
	The Berkeley Proxy Kernel \cite{ref-chipyard} proxies single-threaded syscalls via the host–target interface without booting a full OS, but it remains simulator-bound and requires a complete SoC implementation.
	To our knowledge, multi-core syscall emulation on RTL-modeled hardware has not been previously explored.

\subsection{Early-Stage Processor Prototype Evaluation}


	

	Prior work improves early performance evaluation by modeling selected microarchitectural features (e.g., GEM5 accelerator extensions \cite{ref-gem5-accelerator-extension, ref-gem5-hetosoc-sim}) or integrating multi-level simulators such as Gem5-Aladdin \cite{ref-gem5-aladdin}, PARADE \cite{ref-parade}, ERAS \cite{ref-eras}, and BZSim \cite{ref-bzsim}.
	However, simulation efficiency and RTL environmental constraints limit workload realism, often requiring customized or preprocessed applications. As a result, accurately capturing real-world performance remains challenging.
	
	Other efforts enhance RTL simulation efficiency (e.g., JIT compilation \cite{ref-cascade}) or build agile design frameworks using high-level HDLs such as Chisel and BlueSpec \cite{ref-graphrtl, ref-scgen, ref-matrix}.
	However, how to integrate these emerging paradigms into the complete chip design workflow remains an open research question.

\subsection{End-to-End Evaluation and Workflow Simplification}

	Processor validation frameworks generally either (1) automate SoC integration and software setup \cite{ref-chipyard, ref-agile-soc, ref-openpiton, ref-chipkit}, or (2) generate hardware and toolchains from high-level architecture specifications \cite{ref-end2end-sim-dlworkload, ref-byg, ref-blackparrot}.
	The former relies on reusable IP and predefined bus protocols, often requiring adapter logic and platform-specific adjustments.
	The latter struggles to provide accurate performance estimation and flexible design due to its highly abstract modeling.
	FireSim \cite{ref-firesim} enables agile multicore validation across FPGAs but remains a full-system approach requiring a complete hardware–software stack and cross-FPGA synchronization support.
	
	FASE differs fundamentally by eliminating, rather than automating, much of the SoC integration and software porting effort.
	Through syscall emulation, it bypasses OS bring-up, bus mismatches, and FPGA portability issues, and can also integrate into existing flows to provide earlier and faster validation with sufficient accuracy.

\section{Conclusion}
\label{section-conclusion}

	In this paper, we presented FASE (FPGA-Assisted Syscall Emulation), the first framework enabling accurate syscall emulation on FPGA for rapid processor performance validation without requiring a full SoC integration.
	FASE combines a minimal, microarchitecture-independent CPU interface, a Host-Target Protocol (HTP) for efficient remote syscall handling, and a host-side runtime supporting I/O redirection, thread synchronization, and virtual memory management.
	FASE is microarchitecture-independent and can be readily applied to processors implementing the corresponding instruction set.
	Hardware-assisted optimizations including hardware-futex handling further reduce remote syscall overhead.
	Implemented with the RISC-V SMP processor Rocket on a Xilinx KCU105 FPGA, FASE achieves over 91.5\% performance evaluation accuracy in benchmark scores and over 95.9\% accuracy in user-mode CPU time, while supporting multi-threaded ELF workloads with low overhead.
	With single-thread CoreMark workload, FASE achieves over 99\% performance evaluation accuracy and achieves over 2000× higher efficiency compared to Proxy Kernel through FPGA acceleration.
	Its modular, open-source design provides a practical platform for end-to-end processor performance evaluation and future extensions.
	
	Future work will focus on maintaining and enhancing the FASE framework by expanding its supported application range and optimizing key performance bottleneck, the host–target channel.
	We also aim to integrate FASE with other open-source processor design and verification toolchains to further reduce the effort required to use FASE, providing processor designers and explorers with a more comprehensive and efficient platform for performance validation.

\bibliographystyle{ieeetr}
\bibliography{citation}

@ARTICLE{ref-sim-techs,
	author={Akram, Ayaz and Sawalha, Lina},
	journal={IEEE Access}, 
	title={A Survey of Computer Architecture Simulation Techniques and Tools}, 
	year={2019},
	volume={7},
	number={},
	pages={78120-78145},
	doi={10.1109/ACCESS.2019.2917698}}

@INPROCEEDINGS{ref-asic-prototype,
	author={Brasen, D. and Saucier, G.},
	booktitle={Proceedings Seventh IEEE International Workshop on Rapid System Prototyping. Shortening the Path from Specification to Prototype}, 
	title={ASIC prototyping with reprogrammable implementations of large ASICs}, 
	year={1996},
	volume={},
	number={},
	pages={127-132},
	doi={10.1109/IWRSP.1996.506739}}

@ARTICLE{ref-rvcnn,
	author={Wang, Shihang and Wang, Xingbo and Xu, Zhiyuan and Chen, Bingzhen and Feng, Chenxi and Wang, Qi and Ye, Terry Tao},
	journal={IEEE Transactions on Computers}, 
	title={Optimizing CNN Computation Using RISC-V Custom Instruction Sets for Edge Platforms}, 
	year={2024},
	volume={73},
	number={5},
	pages={1371-1384},
	doi={10.1109/TC.2024.3362060}}

@INPROCEEDINGS{ref-rv-isa-extend,
	author={Gao, Zhanyuan and Zhao, Laiping and Chen, Haonan},
	booktitle={2022 IEEE/ACIS 22nd International Conference on Computer and Information Science (ICIS)}, 
	title={A Trigonometric Function Instruction Set Extension Method Based on RISC-V}, 
	year={2022},
	volume={},
	number={},
	pages={119-126},
	doi={10.1109/ICIS54925.2022.9882453}}

@inproceedings{ref-cascade,
	author = {Schkufza, Eric and Wei, Michael and Rossbach, Christopher J.},
	title = {Just-In-Time Compilation for Verilog: A New Technique for Improving the FPGA Programming Experience},
	year = {2019},
	isbn = {9781450362405},
	publisher = {Association for Computing Machinery},
	address = {New York, NY, USA},
	url = {https://doi.org/10.1145/3297858.3304010},
	doi = {10.1145/3297858.3304010},
	booktitle = {Proceedings of the Twenty-Fourth International Conference on Architectural Support for Programming Languages and Operating Systems},
	pages = {271–286},
	numpages = {16},
	location = {Providence, RI, USA},
	series = {ASPLOS '19}
}

@INPROCEEDINGS{ref-roofline-e2e,
	author={Ding, Nan and Austin, Brian and Liu, Yang and Mehta, Neil and Farrell, Steven and Blaschke, Johannes P. and Oliker, Leonid and Nam, Hai Ah and Wright, Nicholas J. and Williams, Samuel},
	booktitle={SC24: International Conference for High Performance Computing, Networking, Storage and Analysis}, 
	title={A Workflow Roofline Model for End-to-End Workflow Performance Analysis}, 
	year={2024},
	volume={},
	number={},
	pages={1-15},
	doi={10.1109/SC41406.2024.00071}}

@INPROCEEDINGS{ref-roofline-gpu,
	author={Prashanth, H C and Rao, Madhav},
	booktitle={2024 25th International Symposium on Quality Electronic Design (ISQED)}, 
	title={Roofline Performance Analysis of DNN Architectures on CPU and GPU Systems}, 
	year={2024},
	volume={},
	number={},
	pages={1-8},
	doi={10.1109/ISQED60706.2024.10528752}}

@ARTICLE{ref-cad-assisted-sim1,
	author={Kim, Jinwoo and Murali, Gauthaman and Park, Heechun and Qin, Eric and Kwon, Hyoukjun and Chekuri, Venkata Chaitanya Krishna and Rahman, Nael Mizanur and Dasari, Nihar and Singh, Arvind and Lee, Minah and Torun, Hakki Mert and Roy, Kallol and Swaminathan, Madhavan and Mukhopadhyay, Saibal and Krishna, Tushar and Lim, Sung Kyu},
	journal={IEEE Transactions on Very Large Scale Integration (VLSI) Systems}, 
	title={Architecture, Chip, and Package Codesign Flow for Interposer-Based 2.5-D Chiplet Integration Enabling Heterogeneous IP Reuse}, 
	year={2020},
	volume={28},
	number={11},
	pages={2424-2437},
	doi={10.1109/TVLSI.2020.3015494}}

@INPROCEEDINGS{ref-cad-assisted-sim2,
	author={Kabir, MD Arafat and Peng, Yarui},
	booktitle={2020 25th Asia and South Pacific Design Automation Conference (ASP-DAC)}, 
	title={Chiplet-Package Co-Design For 2.5D Systems Using Standard ASIC CAD Tools}, 
	year={2020},
	volume={},
	number={},
	pages={351-356},
	doi={10.1109/ASP-DAC47756.2020.9045734}}

@ARTICLE{ref-hlpow,
	author={Lin, Zhe and Liang, Tingyuan and Zhao, Jieru and Sinha, Sharad and Zhang, Wei},
	journal={IEEE Transactions on Computer-Aided Design of Integrated Circuits and Systems}, 
	title={HL-Pow: Learning-Assisted Pre-RTL Power Modeling and Optimization for FPGA HLS}, 
	year={2023},
	volume={42},
	number={11},
	pages={3925-3938},
	doi={10.1109/TCAD.2023.3246387}}

@INPROCEEDINGS{ref-gem5-aladdin,
	author={Shao, Yakun Sophia and Xi, Sam Likun and Srinivasan, Vijayalakshmi and Wei, Gu-Yeon and Brooks, David},
	booktitle={2016 49th Annual IEEE/ACM International Symposium on Microarchitecture (MICRO)}, 
	title={Co-designing accelerators and SoC interfaces using gem5-Aladdin}, 
	year={2016},
	volume={},
	number={},
	pages={1-12},
	doi={10.1109/MICRO.2016.7783751}}

@INPROCEEDINGS{ref-parade,
	author={Cong, Jason and Fang, Zhenman and Gill, Michael and Reinman, Glenn},
	booktitle={2015 IEEE/ACM International Conference on Computer-Aided Design (ICCAD)}, 
	title={PARADE: A cycle-accurate full-system simulation Platform for Accelerator-Rich Architectural Design and Exploration}, 
	year={2015},
	volume={},
	number={},
	pages={380-387},
	doi={10.1109/ICCAD.2015.7372595}}

@INPROCEEDINGS{ref-eras,
	author={Nema, Shubham and Chunduru, Shiva Kaushik and Kodigal, Charan and Voskuilen, Gwendolyn and Rodrigues, Arun F. and Hemmert, Scott and Feinberg, Ben and Lee, Hyokeun and Awad, Amro and Hughes, Clayton},
	booktitle={2023 IEEE International Symposium on Workload Characterization (IISWC)}, 
	title={ERAS: A Flexible and Scalable Framework for Seamless Integration of RTL Models with Structural Simulation Toolkit}, 
	year={2023},
	volume={},
	number={},
	pages={196-200},
	doi={10.1109/IISWC59245.2023.00038}}

@INPROCEEDINGS{ref-bzsim,
	author={Strikos, Panagiotis and Ejaz, Ahsen and Sourdis, Ioannis},
	booktitle={2024 IEEE International Symposium on Performance Analysis of Systems and Software (ISPASS)}, 
	title={BZSim: Fast, Large-Scale Microarchitectural Simulation with Detailed Interconnect Modeling}, 
	year={2024},
	volume={},
	number={},
	pages={167-178},
	doi={10.1109/ISPASS61541.2024.00025}}

@INPROCEEDINGS{ref-graphrtl,
	author={Qiao, Yuheng and Xie, Cai and Ou, Zhaoting and Lei, Peizhi and Tian, Yan and Chen, Jienan},
	booktitle={2024 2nd International Symposium of Electronics Design Automation (ISEDA)}, 
	title={GraphRTL: An Agile Design Framework of RTL Code from Data Flow Graphs}, 
	year={2024},
	volume={},
	number={},
	pages={259-265},
	doi={10.1109/ISEDA62518.2024.10617508}}

@INPROCEEDINGS{ref-scgen,
	author={Li, Zexi and Jin, Haoran and Zhong, Kuncai and Luo, Guojie and Wang, Runsheng and Qian, Weikang},
	booktitle={2024 Design, Automation \& Test in Europe Conference \& Exhibition (DATE)}, 
	title={SCGen: A Versatile Generator Framework for Agile Design of Stochastic Circuits}, 
	year={2024},
	volume={},
	number={},
	pages={1-6},
	doi={10.23919/DATE58400.2024.10546649}}

@INPROCEEDINGS{ref-matrix,
	author={Li, Jiangnan and Yan, Yazhou and Zhu, Guowei and Yin, Wenbo and Wang, Lingli},
	booktitle={2024 IEEE 35th International Conference on Application-specific Systems, Architectures and Processors (ASAP)}, 
	title={An End-to-End Agile Design Framework to Improve Energy Efficiency on CGRAs}, 
	year={2024},
	volume={},
	number={},
	pages={17-18},
	doi={10.1109/ASAP61560.2024.00014}}

@ARTICLE{ref-chipkit,
	author={Whatmough, Paul N. and Donato, Marco and Ko, Glenn G. and Lee, Sae Kyu and Brooks, David and Wei, Gu-Yeon},
	journal={IEEE Micro}, 
	title={CHIPKIT: An Agile, Reusable Open-Source Framework for Rapid Test Chip Development}, 
	year={2020},
	volume={40},
	number={4},
	pages={32-40},
	doi={10.1109/MM.2020.2995809}}

@ARTICLE{ref-modse-exploration,
	author={Wang, Duo and Yan, Mingyu and Teng, Yihan and Han, Dengke and Liu, Xin and Li, Wenming and Ye, Xiaochun and Fan, Dongrui},
	journal={IEEE Transactions on Computer-Aided Design of Integrated Circuits and Systems}, 
	title={MoDSE: A High-Accurate Multiobjective Design Space Exploration Framework for CPU Microarchitectures}, 
	year={2024},
	volume={43},
	number={5},
	pages={1525-1537},
	doi={10.1109/TCAD.2023.3340059}}

@article{ref-openpiton,
	author = {Balkind, Jonathan and McKeown, Michael and Fu, Yaosheng and Nguyen, Tri and Zhou, Yanqi and Lavrov, Alexey and Shahrad, Mohammad and Fuchs, Adi and Payne, Samuel and Liang, Xiaohua and Matl, Matthew and Wentzlaff, David},
	title = {OpenPiton: An Open Source Manycore Research Framework},
	year = {2016},
	issue_date = {May 2016},
	publisher = {Association for Computing Machinery},
	address = {New York, NY, USA},
	volume = {44},
	number = {2},
	issn = {0163-5964},
	url = {https://doi.org/10.1145/2980024.2872414},
	doi = {10.1145/2980024.2872414},
	journal = {SIGARCH Comput. Archit. News},
	month = mar,
	pages = {217–232},
	numpages = {16}
}

@ARTICLE{ref-blackparrot,
	author={Petrisko, Daniel and Gilani, Farzam and Wyse, Mark and Jung, Dai Cheol and Davidson, Scott and Gao, Paul and Zhao, Chun and Azad, Zahra and Canakci, Sadullah and Veluri, Bandhav and Guarino, Tavio and Joshi, Ajay and Oskin, Mark and Taylor, Michael Bedford},
	journal={IEEE Micro}, 
	title={BlackParrot: An Agile Open-Source RISC-V Multicore for Accelerator SoCs}, 
	year={2020},
	volume={40},
	number={4},
	pages={93-102},
	doi={10.1109/MM.2020.2996145}}

@INPROCEEDINGS{ref-byg,
	author={Xu, Yinan and Yu, Zihao and Tang, Dan and Chen, Guokai and Chen, Lu and Gou, Lingrui and Jin, Yue and Li, Qianruo and Li, Xin and Li, Zuojun and Lin, Jiawei and Liu, Tong and Liu, Zhigang and Tan, Jiazhan and Wang, Huaqiang and Wang, Huizhe and Wang, Kaifan and Zhang, Chuanqi and Zhang, Fawang and Zhang, Linjuan and Zhang, Zifei and Zhao, Yangyang and Zhou, Yaoyang and Zhou, Yike and Zou, Jiangrui and Cai, Ye and Huan, Dandan and Li, Zusong and Zhao, Jiye and Chen, Zihao and He, Wei and Quan, Qiyuan and Liu, Xingwu and Wang, Sa and Shi, Kan and Sun, Ninghui and Bao, Yungang},
	booktitle={2022 55th IEEE/ACM International Symposium on Microarchitecture (MICRO)}, 
	title={Towards Developing High Performance RISC-V Processors Using Agile Methodology}, 
	year={2022},
	volume={},
	number={},
	pages={1178-1199},
	doi={10.1109/MICRO56248.2022.00080}}

@ARTICLE{ref-chipyard,
	author={Amid, Alon and Biancolin, David and Gonzalez, Abraham and Grubb, Daniel and Karandikar, Sagar and Liew, Harrison and Magyar, Albert and Mao, Howard and Ou, Albert and Pemberton, Nathan and Rigge, Paul and Schmidt, Colin and Wright, John and Zhao, Jerry and Shao, Yakun Sophia and Asanović, Krste and Nikolić, Borivoje},
	journal={IEEE Micro}, 
	title={Chipyard: Integrated Design, Simulation, and Implementation Framework for Custom SoCs}, 
	year={2020},
	volume={40},
	number={4},
	pages={10-21},
	doi={10.1109/MM.2020.2996616}}

@INPROCEEDINGS{ref-agile-soc,
	author={Mantovani, Paolo and Giri, Davide and Di Guglielmo, Giuseppe and Piccolboni, Luca and Zuckerman, Joseph and Cota, Emilio G. and Petracca, Michele and Pilato, Christian and Carloni, Luca P.},
	booktitle={2020 IEEE/ACM International Conference On Computer Aided Design (ICCAD)}, 
	title={Agile SoC Development with Open ESP}, 
	year={2020},
	volume={},
	number={},
	pages={1-9},
	doi={}}

@INPROCEEDINGS{ref-presi,
	author={Fadiheh, Mohammad Rahmani and Urdahl, Joakim and Nuthakki, Srinivas Shashank and Mitra, Subhasish and Barrett, Clark and Stoffel, Dominik and Kunz, Wolfgang},
	booktitle={2018 Design, Automation \& Test in Europe Conference \& Exhibition (DATE)}, 
	title={Symbolic quick error detection using symbolic initial state for pre-silicon verification}, 
	year={2018},
	volume={},
	number={},
	pages={55-60},
	doi={10.23919/DATE.2018.8341979}}

@ARTICLE{ref-agile-test,
	author={Whatmough, Paul N. and Donato, Marco and Ko, Glenn G. and Lee, Sae Kyu and Brooks, David and Wei, Gu-Yeon},
	journal={IEEE Micro}, 
	title={CHIPKIT: An Agile, Reusable Open-Source Framework for Rapid Test Chip Development}, 
	year={2020},
	volume={40},
	number={4},
	pages={32-40},
	doi={10.1109/MM.2020.2995809}}

@INPROCEEDINGS{ref-fireaxe-large-rtl-sim,
	author={Whangbo, Joonho and Lim, Edwin and Zhang, Chengyi Lux and Anderson, Kevin and Gonzalez, Abraham and Gupta, Raghav and Krishnakumar, Nivedha and Karandikar, Sagar and Nikolić, Borivoje and Shao, Yakun Sophia and Asanović, Krste},
	booktitle={2024 ACM/IEEE 51st Annual International Symposium on Computer Architecture (ISCA)}, 
	title={FireAxe: Partitioned FPGA-Accelerated Simulation of Large-Scale RTL Designs}, 
	year={2024},
	volume={},
	number={},
	pages={501-515},
	doi={10.1109/ISCA59077.2024.00044}}

@INPROCEEDINGS{ref-firesim,
  author={Karandikar, Sagar and Mao, Howard and Kim, Donggyu and Biancolin, David and Amid, Alon and Lee, Dayeol and Pemberton, Nathan and Amaro, Emmanuel and Schmidt, Colin and Chopra, Aditya and Huang, Qijing and Kovacs, Kyle and Nikolic, Borivoje and Katz, Randy and Bachrach, Jonathan and Asanovic, Krste},
  booktitle={2018 ACM/IEEE 45th Annual International Symposium on Computer Architecture (ISCA)}, 
  title={FireSim: FPGA-Accelerated Cycle-Exact Scale-Out System Simulation in the Public Cloud}, 
  year={2018},
  volume={},
  number={},
  pages={29-42},
  doi={10.1109/ISCA.2018.00014}}

@INPROCEEDINGS{ref-gem5-hetosoc-sim,
	author={Chatzopoulos, Odysseas and Papadimitriou, George and Karakostas, Vasileios and Gizopoulos, Dimitris},
	booktitle={2024 IEEE International Symposium on High-Performance Computer Architecture (HPCA)}, 
	title={Gem5-MARVEL: Microarchitecture-Level Resilience Analysis of Heterogeneous SoC Architectures}, 
	year={2024},
	volume={},
	number={},
	pages={543-559},
	doi={10.1109/HPCA57654.2024.00047}}

@INPROCEEDINGS{ref-hwsw-rtl-sim,
	author={Elsabbagh, Fares and Sheikhha, Shabnam and Ying, Victor A. and Nguyen, Quan M. and Emer, Joel S. and Sanchez, Daniel},
	booktitle={2023 56th IEEE/ACM International Symposium on Microarchitecture (MICRO)}, 
	title={Accelerating RTL Simulation with Hardware-Software Co-Design}, 
	year={2023},
	volume={},
	number={},
	pages={153-166},
	doi={}}

@INPROCEEDINGS{ref-gem5-accelerator-extension,
	author={Rogers, Samuel and Slycord, Joshua and Baharani, Mohammadreza and Tabkhi, Hamed},
	booktitle={2020 53rd Annual IEEE/ACM International Symposium on Microarchitecture (MICRO)}, 
	title={gem5-SALAM: A System Architecture for LLVM-based Accelerator Modeling}, 
	year={2020},
	volume={},
	number={},
	pages={471-482},
	doi={10.1109/MICRO50266.2020.00047}}

@article{ref-manycore-framework,
	author = {Ruaro, Marcelo and Sant’ana, Anderson and Jantsch, Axel and Moraes, Fernando Gehm},
	title = {Modular and Distributed Management of Many-Core SoCs},
	year = {2021},
	issue_date = {May 2020},
	publisher = {Association for Computing Machinery},
	address = {New York, NY, USA},
	volume = {38},
	number = {1–2},
	issn = {0734-2071},
	url = {https://doi.org/10.1145/3458511},
	doi = {10.1145/3458511},
	journal = {ACM Trans. Comput. Syst.},
	month = jul,
	articleno = {1},
	numpages = {16}
}

@article{ref-rvvec-sim-bench-nogem5,
	author = {Ram\'{\i}rez, Crist\'{o}bal and Hern\'{a}ndez, C\'{e}sar Alejandro and Palomar, Oscar and Unsal, Osman and Ram\'{\i}rez, Marco Antonio and Cristal, Adri\'{a}n},
	title = {A RISC-V Simulator and Benchmark Suite for Designing and Evaluating Vector Architectures},
	year = {2020},
	issue_date = {December 2020},
	publisher = {Association for Computing Machinery},
	address = {New York, NY, USA},
	volume = {17},
	number = {4},
	issn = {1544-3566},
	url = {https://doi.org/10.1145/3422667},
	doi = {10.1145/3422667},
	journal = {ACM Trans. Archit. Code Optim.},
	month = nov,
	articleno = {38},
	numpages = {30}
}

@article{ref-survey-cachesim,
	author = {Brais, Hadi and Kalayappan, Rajshekar and Panda, Preeti Ranjan},
	title = {A Survey of Cache Simulators},
	year = {2020},
	issue_date = {January 2021},
	publisher = {Association for Computing Machinery},
	address = {New York, NY, USA},
	volume = {53},
	number = {1},
	issn = {0360-0300},
	url = {https://doi.org/10.1145/3372393},
	doi = {10.1145/3372393},
	journal = {ACM Comput. Surv.},
	month = feb,
	articleno = {19},
	numpages = {32}
}

@article{ref-end2end-sim-dlworkload,
	author = {Xi, Sam (Likun) and Yao, Yuan and Bhardwaj, Kshitij and Whatmough, Paul and Wei, Gu-Yeon and Brooks, David},
	title = {SMAUG: End-to-End Full-Stack Simulation Infrastructure for Deep Learning Workloads},
	year = {2020},
	issue_date = {December 2020},
	publisher = {Association for Computing Machinery},
	address = {New York, NY, USA},
	volume = {17},
	number = {4},
	issn = {1544-3566},
	url = {https://doi.org/10.1145/3424669},
	doi = {10.1145/3424669},
	journal = {ACM Trans. Archit. Code Optim.},
	month = nov,
	articleno = {39},
	numpages = {26}
}

@article{ref-gem5,
	author = {Binkert, Nathan and Beckmann, Bradford and Black, Gabriel and Reinhardt, Steven K. and Saidi, Ali and Basu, Arkaprava and Hestness, Joel and Hower, Derek R. and Krishna, Tushar and Sardashti, Somayeh and Sen, Rathijit and Sewell, Korey and Shoaib, Muhammad and Vaish, Nilay and Hill, Mark D. and Wood, David A.},
	title = {The gem5 simulator},
	year = {2011},
	issue_date = {May 2011},
	publisher = {Association for Computing Machinery},
	address = {New York, NY, USA},
	volume = {39},
	number = {2},
	issn = {0163-5964},
	url = {https://doi.org/10.1145/2024716.2024718},
	doi = {10.1145/2024716.2024718},
	journal = {SIGARCH Comput. Archit. News},
	month = aug,
	pages = {1–7},
	numpages = {7}
}

@article{ref-zsim,
	author = {Sanchez, Daniel and Kozyrakis, Christos},
	title = {ZSim: fast and accurate microarchitectural simulation of thousand-core systems},
	year = {2013},
	issue_date = {June 2013},
	publisher = {Association for Computing Machinery},
	address = {New York, NY, USA},
	volume = {41},
	number = {3},
	issn = {0163-5964},
	url = {https://doi.org/10.1145/2508148.2485963},
	doi = {10.1145/2508148.2485963},
	journal = {SIGARCH Comput. Archit. News},
	month = jun,
	pages = {475–486},
	numpages = {12}
}

@INPROCEEDINGS{ref-sniper,
	author={Carlson, Trevor E. and Heirman, Wim and Eeckhout, Lieven},
	booktitle={SC '11: Proceedings of 2011 International Conference for High Performance Computing, Networking, Storage and Analysis}, 
	title={Sniper: Exploring the level of abstraction for scalable and accurate parallel multi-core simulation}, 
	year={2011},
	volume={},
	number={},
	pages={1-12},
	doi={10.1145/2063384.2063454}}

@INPROCEEDINGS{ref-mcsima,
	author={Ahn, Jung Ho and Li, Sheng and O, Seongil and Jouppi, Norman P.},
	booktitle={2013 IEEE International Symposium on Performance Analysis of Systems and Software (ISPASS)}, 
	title={McSimA+: A manycore simulator with application-level+ simulation and detailed microarchitecture modeling}, 
	year={2013},
	volume={},
	number={},
	pages={74-85},
	doi={10.1109/ISPASS.2013.6557148}}

@INPROCEEDINGS{ref-esesc,
	author={Ardestani, Ehsan K. and Renau, Jose},
	booktitle={2013 IEEE 19th International Symposium on High Performance Computer Architecture (HPCA)}, 
	title={ESESC: A fast multicore simulator using Time-Based Sampling}, 
	year={2013},
	volume={},
	number={},
	pages={448-459},
	doi={10.1109/HPCA.2013.6522340}}

@INPROCEEDINGS{ref-vmsim,
	author={Tchana, Alain and Ekane, Brice and Teabe, Boris and Hagimont, Daniel},
	booktitle={2015 IEEE 8th International Conference on Cloud Computing}, 
	title={VMcSim: A Detailed Manycore Simulator for Virtualized Systems}, 
	year={2015},
	volume={},
	number={},
	pages={195-202},
	doi={10.1109/CLOUD.2015.35}}

@INPROCEEDINGS{ref-boom-sim,
	author={Bai, Chen and Sun, Qi and Zhai, Jianwang and Ma, Yuzhe and Yu, Bei and Wong, Martin D.F.},
	booktitle={2021 IEEE/ACM International Conference On Computer Aided Design (ICCAD)}, 
	title={BOOM-Explorer: RISC-V BOOM Microarchitecture Design Space Exploration Framework}, 
	year={2021},
	volume={},
	number={},
	pages={1-9},
	doi={10.1109/ICCAD51958.2021.9643455}}

\end{document}